\documentclass[aps,prd,superscriptaddress,nofootinbib,twocolumn,preprintnumbers]{revtex4-2}

\usepackage{amsmath,amssymb,amsfonts,bm,slashed,graphicx,array,multirow,xspace}
\usepackage{placeins,comment}
\usepackage{setspace}
\usepackage{booktabs}
\usepackage{makecell}
\usepackage{xstring}
\usepackage[labelfont=bf,font={small,stretch=1},
  justification=centerlast]{caption}
\usepackage[shortlabels]{enumitem}
\usepackage[bookmarks=false]{hyperref}
\usepackage[dvipsnames]{xcolor}
\usepackage{cleveref}
\usepackage{orcidlink}
\usepackage{makecell}
\usepackage{natbib} 
\usepackage{bibentry}
\nobibliography*

\DeclareRobustCommand{\ccite}[1]{\IfSubStr{#1}{,}{refs.~}{ref.~}\cite{#1}}
\DeclareRobustCommand{\Ccite}[1]{\IfSubStr{#1}{,}{Refs.~}{Ref.~}\cite{#1}}

\newcommand{\eg}{\textit{e.g.},\xspace}
\newcommand{\ie}{\textit{i.e.},\xspace}

\newcommand{\cbsA}{baseline\xspace}
\newcommand{\cbsB}{scenario~1\xspace}
\newcommand{\cbsC}{scenario~2\xspace}
\newcommand{\cbsD}{scenario~3\xspace}

\newcommand{\CbsB}{Scenario~1\xspace}
\newcommand{\CbsC}{Scenario~2\xspace}

\newcommand{\cm}{\text{cm}}
\newcommand{\gev}{\text{GeV}}
\newcommand{\mev}{\text{MeV}}
\newcommand{\CODEXb}{\mbox{CODEX-b}\xspace}
\newcommand{\CODEXbeta}{\mbox{CODEX-$\beta$}\xspace}

\makeatletter
\g@addto@macro\bfseries{\boldmath}
\makeatother

\definecolor{nicered}{rgb}{0.7,0.1,0.1}
\definecolor{nicegreen}{rgb}{0.1,0.5,0.1}
\definecolor{niceblue}{rgb}{0.1,0.1,0.7}
\hypersetup{colorlinks,citecolor= nicegreen,linkcolor= niceblue}

\allowdisplaybreaks
\makeatletter
\let\temp@subsection\subsection
\renewcommand{\subsection}[1]{\temp@subsection{#1}\vspace{-10pt}}
\makeatother

\begin{document}

\preprint{\includegraphics[width=3cm]{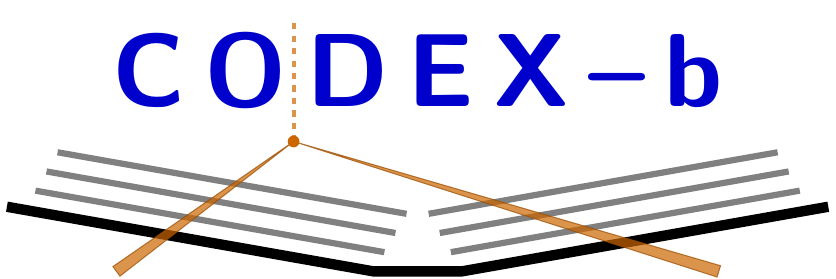}}
\title{CODEX-b: Opening New Windows to the Long-Lived Particle Frontier at the LHC\\ {\normalsize \normalfont ESPP Update} \\ {\normalsize  Comprehensive summary}}

\author{Giulio Aielli\,\orcidlink{0000-0002-0573-8114}}
\affiliation{Universit\`a e INFN Sezione di Roma Tor Vergata, Roma, Italy}

\author{Juliette Alimena\,\orcidlink{0000-0001-6030-3191}}
\affiliation{Deutsches Elektronen-Synchrotron (DESY), Hamburg, Germany}

\author{Saul Balcarcel-Salazar\,\orcidlink{0009-0005-1104-3502}}
\affiliation{Laboratory for Nuclear Science, Massachusetts Institute of Technology, Cambridge, MA 02139, USA}

\author{Eli Ben Haim\,\orcidlink{0000-0002-9510-8414}}
\affiliation{LPNHE, Sorbonne Universit\'e, Universit{\'e} Paris Cit{\'e}, CNRS/IN2P3, Paris, France}

\author{Andr{\'a}s Barnab{\'a}s Burucs\,\orcidlink{009-0001-6791-1007}}
\affiliation{ELTE E\"otv\"os Lor\'and University, Budapest, Hungary}

\author{Roberto Cardarelli\,\orcidlink{0000-0003-4541-4189}}
\affiliation{Universit\`a e INFN Sezione di Roma Tor Vergata, Roma, Italy}

\author{Matthew J.~Charles\,\orcidlink{0000-0003-4795-498X}}
\affiliation{LPNHE, Sorbonne Universit\'e, Universit{\'e} Paris Cit{\'e}, CNRS/IN2P3, Paris, France}

\author{Xabier Cid Vidal\,\orcidlink{0000-0002-0468-541X}}
\affiliation{Instituto Galego de F\'isica de Altas Enerx\'ias (IGFAE), Universidade de Santiago de Compostela, Santiago de Compostela, Spain}

\author{Albert De Roeck\,\orcidlink{0000-0002-9228-5271}}
\affiliation{European Organization for Nuclear Research (CERN), Geneva, Switzerland}

\author{Biplab Dey\,\orcidlink{0000-0002-4563-5806}}
\affiliation{ELTE E\"otv\"os Lor\'and University, Budapest, Hungary}

\author{Silviu Dobrescu}
\affiliation{Department of Physics, University of Cincinnati, Cincinnati, OH 45221, USA}

\author{Ozgur Durmus\,\orcidlink{0000-0002-8161-7832}}
\affiliation{ELTE E\"otv\"os Lor\'and University, Budapest, Hungary}

\author{Mohamed Elashri\,\orcidlink{0000-0001-9398-953X}}
\affiliation{Department of Physics, University of Cincinnati, Cincinnati, OH 45221, USA}

\author{Vladimir V.~Gligorov\,\orcidlink{0000-0002-8189-8267}}
\affiliation{LPNHE, Sorbonne Universit\'e, Universit{\'e} Paris Cit{\'e}, CNRS/IN2P3, Paris, France}

\author{Rebeca Gonzalez Suarez\,\orcidlink{0000-0002-6126-7230}}
\affiliation{Department of Physics and Astronomy, Uppsala Universitet, Uppsala, Sweden}

\author{Zarria Gray}
\affiliation{Department of Physics, University of Cincinnati, Cincinnati, OH 45221, USA}

\author{Conor Henderson\,\orcidlink{0000-0002-6986-9404}}
\affiliation{Department of Physics, University of Cincinnati, Cincinnati, OH 45221, USA}

\author{Louis Henry\,\orcidlink{0000-0003-3605-832X}}
\affiliation{Ecole Polytechnique F\'ed\'erale Lausanne, Lausanne, Switzerland}

\author{Philip Ilten\,\orcidlink{0000-0001-5534-1732}}
\affiliation{Department of Physics, University of Cincinnati, Cincinnati, OH 45221, USA}

\author{Daniel Johnson\,\orcidlink{0000-0003-3272-6001}}
\affiliation{University of Birmingham, Birmingham, United Kingdom}

\author{Jacob Kautz\,\orcidlink{0000-0001-8482-5576}}
\affiliation{Department of Physics, University of Cincinnati, Cincinnati, OH 45221, USA}

\author{Simon Knapen\,\orcidlink{0000-0002-6733-9231}}
\affiliation{Physics Division, Lawrence Berkeley National Laboratory, Berkeley, CA 94720, USA}
\affiliation{Department of Physics, University of California, Berkeley, CA 94720, USA}

\author{Bingxuan Liu\,\orcidlink{0000-0002-0721-8331}}
\affiliation{School of Science, Shenzhen Campus of Sun Yat-Sen University, Shenzhen, China}

\author{Yang Liu\,\orcidlink{0000-0003-3615-2332}}
\affiliation{School of Science, Shenzhen Campus of Sun Yat-Sen University, Shenzhen, China}

\author{Saul L\'opez Soli\~no\,\orcidlink{0000-0001-9892-5113}}
\affiliation{Instituto Galego de F\'isica de Altas Enerx\'ias (IGFAE), Universidade de Santiago de Compostela, Santiago de Compostela, Spain}

\author{Pablo Eduardo Men\'endez-Vald\'es P\'erez\,\orcidlink{0009-0003-0406-8141}}
\affiliation{Instituto Galego de F\'isica de Altas Enerx\'ias (IGFAE), Universidade de Santiago de Compostela, Santiago de Compostela, Spain}

\author{Titus Momb\"acher\,\orcidlink{0000-0002-5612-979X}}
\affiliation{European Organization for Nuclear Research (CERN), Geneva, Switzerland}

\author{Benjamin Nachman\,\orcidlink{0000-0003-1024-0932}}
\affiliation{Physics Division, Lawrence Berkeley National Laboratory, Berkeley, CA 94720, USA}

\author{David T.~Northacker}
\affiliation{Department of Physics, University of Cincinnati, Cincinnati, OH 45221, USA}

\author{Gabriel M.~Nowak\,\orcidlink{0000-0003-4864-7164}}
\affiliation{Department of Physics, University of Cincinnati, Cincinnati, OH 45221, USA}

\author{Michele Papucci\,\orcidlink{0000-0003-0810-0017}}
\affiliation{Walter Burke Institute for Theoretical Physics, California Institute of Technology, Pasadena, CA 91125, USA}

\author{Gabriella P\'asztor\,\orcidlink{0000-0003-0707-9762}}
\affiliation{ELTE E\"otv\"os Lor\'and University, Budapest, Hungary}

\author{Mar\'ia Pereira Mart\'inez\,\orcidlink{0009-0006-8577-9560}}
\affiliation{Instituto Galego de F\'isica de Altas Enerx\'ias (IGFAE), Universidade de Santiago de Compostela, Santiago de Compostela, Spain}

\author{Michael Peters\,\orcidlink{0009-0008-9089-1287}}
\affiliation{Department of Physics, University of Cincinnati, Cincinnati, OH 45221, USA}

\author{Jake Pfaller\,\orcidlink{0009-0009-8578-3078}}
\affiliation{Department of Physics, University of Cincinnati, Cincinnati, OH 45221, USA}

\author{Luca Pizzimento\,\orcidlink{0000-0002-1814-2758}}
\affiliation{University of Hong Kong, Hong Kong}

\author{M\'aximo Pl\'o Casas\'us\,\orcidlink{0000-0002-2289-918X}}
\affiliation{Instituto Galego de F\'isica de Altas Enerx\'ias (IGFAE), Universidade de Santiago de Compostela, Santiago de Compostela, Spain}

\author{Gian Andrea Rassati\,\orcidlink{0000-0002-7782-2155}}
\affiliation{Department of Physics, University of Cincinnati, Cincinnati, OH 45221, USA}

\author{Dean J.~Robinson\,\orcidlink{0000-0003-0057-1703}}
\affiliation{Physics Division, Lawrence Berkeley National Laboratory, Berkeley, CA 94720, USA}
\affiliation{Department of Physics, University of California, Berkeley, CA 94720, USA}

\author{Emilio Xos\'e Rodr\'iguez Fern\'andez\,\orcidlink{0000-0002-3040-065X}}
\affiliation{Instituto Galego de F\'isica de Altas Enerx\'ias (IGFAE), Universidade de Santiago de Compostela, Santiago de Compostela, Spain}

\author{Debashis Sahoo\,\orcidlink{0000-0002-5600-9413}}
\affiliation{ELTE E\"otv\"os Lor\'and University, Budapest, Hungary}

\author{Sinem Simsek\,\orcidlink{0000-0002-9650-3846}}
\affiliation{Istynie University, Istanbul, Turkey}

\author{Michael D.~Sokoloff\,\orcidlink{0000-0002-1468-0479}}
\affiliation{Department of Physics, University of Cincinnati, Cincinnati, OH 45221, USA}

\author{Joeal Subash\,\orcidlink{0000-0001-6431-6010}}
\affiliation{University of Birmingham, Birmingham, United Kingdom}

\author{James Swanson\,\orcidlink{0000-0002-5501-3867}}
\affiliation{Department of Physics, University of Cincinnati, Cincinnati, OH 45221, USA}

\author{Abhinaba Upadhyay\,\orcidlink{0009-0000-6052-6889}}
\affiliation{ELTE E\"otv\"os Lor\'and University, Budapest, Hungary}

\author{Riccardo Vari\,\orcidlink{0000-0002-2814-1337}}
\affiliation{INFN Sezione di Roma La Sapienza, Roma, Italy}

\author{Carlos V\'azquez~Sierra\,\orcidlink{0000-0002-5865-0677}}
\affiliation{Universidade da Coru\~na, A Coru\~na, Galicia, Spain}

\author{G\'abor Veres\,\orcidlink{0000-0002-5440-4356}}
\affiliation{ELTE E\"otv\"os Lor\'and University, Budapest, Hungary}

\author{Nigel Watson\,\orcidlink{0000-0002-8142-4678}}
\affiliation{University of Birmingham, Birmingham, United Kingdom}

\author{Michael K.~Wilkinson\,\orcidlink{0000-0001-6561-2145}}
\affiliation{Department of Physics, University of Cincinnati, Cincinnati, OH 45221, USA}

\author{Michael Williams\,\orcidlink{0000-0001-8285-3346}}
\affiliation{Laboratory for Nuclear Science, Massachusetts Institute of Technology, Cambridge, MA 02139, USA}

\author{Eleanor Winkler\,\orcidlink{0000-0002-5021-4002}}
\affiliation{Laboratory for Nuclear Science, Massachusetts Institute of Technology, Cambridge, MA 02139, USA}

\collaboration{CODEX-b collaboration}

\begin{abstract}
This document is written as a contribution to the European Strategy of Particle Physics (ESPP) update.
We offer a detailed overview of current developments and future directions for the \CODEXb detector, which aims to detect long-lived particles beyond the Standard Model.
We summarize the scientific motivation for this detector, advances in our suite of simulation and detector optimization frameworks, and examine expected challenges, costs, and timelines in realizing the full detector.
Additionally, we describe the technical specifications for the smaller-scale demonstrator detector (\CODEXbeta) we have installed in the LHCb experimental cavern.
\end{abstract}
\maketitle
\onecolumngrid

\newpage

\tableofcontents
\thispagestyle{empty}

\newpage
\setcounter{page}{0}
\twocolumngrid
\section{Introduction and objectives}

The primary LHC experiments (ATLAS, CMS, LHCb, ALICE) are scheduled to collect data until at least $2040$.
A core component of the (HL-)LHC program is searches for dark or hidden sectors beyond the Standard Model (BSM), which may be probed in many generic BSM scenarios via searches for displaced decays-in-flight of exotic long-lived particles (LLPs).
In the Standard Model (SM), LLPs such as the $K^0_L$, $\pi^\pm$, neutron and muon arise via suppressions from scale hierarchies ($m_W\gg \Lambda_{\text{QCD}}$),  loop and phase-space suppressions and/or the smallness of one or more CKM matrix elements.  
LLPs can arise similarly and ubiquitously in BSM scenarios featuring \eg Dark Matter, Baryogenesis, Supersymmetry, or Neutral Naturalness.

\begin{figure}
  \centering
  \includegraphics[width=0.45\textwidth]{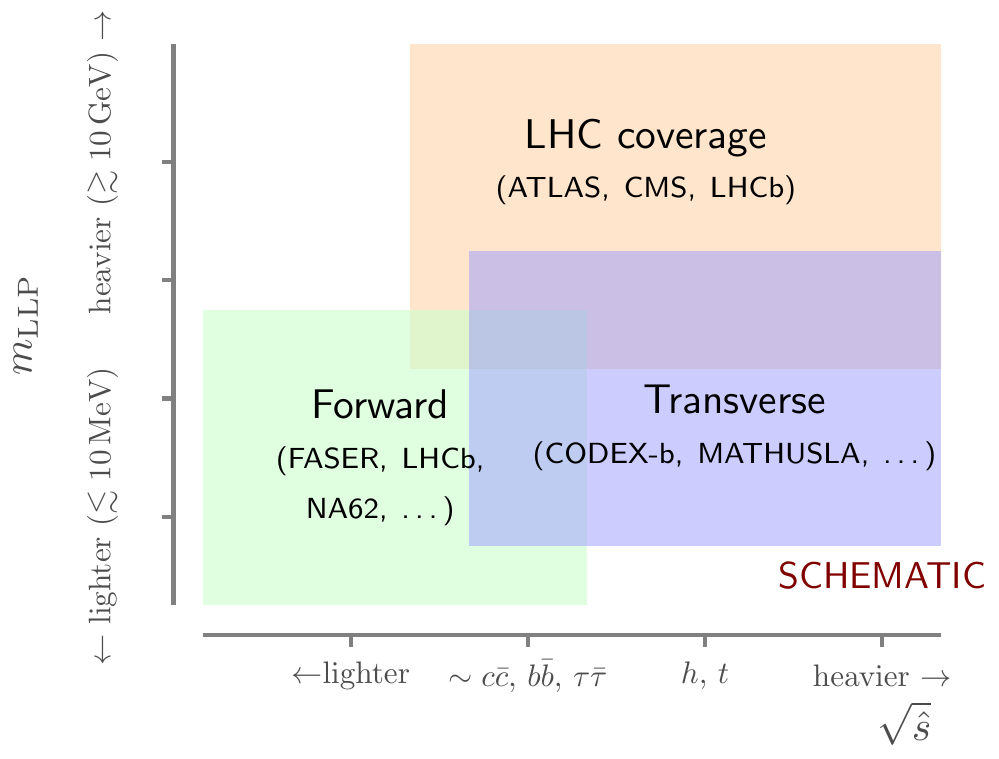} 
  \caption{Complementarity of different experiments searching for LLPs~\cite{Aielli:2019ivi}.\label{fig:schematic}}
\end{figure}

ATLAS, CMS, and LHCb have vibrant LLP search programs that draw on the expertise of analysts, detector specialists, and theorists~\cite{Alimena:2019zri}. 
Typically, both ATLAS and CMS have non-systematics-limited sensitivities for LLPs that are relatively heavy ($m\gtrsim 10~\gev$), while backgrounds and trigger challenges strongly limit the reach for light LLPs due to the complicated environment inherent to a high-energy, high-intensity hadron collider; there are, however, some exceptions, \eg \ccite{CMS:2021sch,CMS:2021juv}.
In the case of LHCb, these difficulties are somewhat offset by its high-precision VErtex LOcator (VELO), but LLP sensitivity is nonetheless restricted to relatively short lifetimes and production at low center-of-mass energies; sensitivity to LLPs produced in, \eg exotic Higgs or $B$ decays can be similarly limited, especially for $c\tau \gtrsim 1\,\mathrm{m}$.
Similar considerations apply to FASER, which leverages a very large amount of passive shielding in the very forward regime.
Thus, to achieve comprehensive coverage of the full LLP parametric landscape, 
one or more high volume, \textit{transverse} LLP detectors are needed.
The complementarity of these transverse detectors is shown in \cref{fig:schematic}.

\textbf{\CODEXb~(\textit{``COmpact Detector for EXotics at LHCb"})} is a special-purpose detector proposed to be installed near and transverse to the LHCb interaction point (IP8) in order to search for displaced decays-in-flight of exotic LLPs~\cite{Gligorov:2017nwh,Aielli:2019ivi,Dey:2019vyo}.
The core advantages of \CODEXb are
\begin{enumerate}[a),noitemsep]
\item competitive sensitivity to a broad array of BSM LLP scenarios, exceeding or complementing the sensitivity of existing or proposed detectors;
\item a zero background environment in an accessible experimental location with many of the necessary services already in place;
\item the possibility to tag events of interest within the LHCb detector (independent of the LHCb physics program); and
\item its compact size and use of existing technology and infrastructure, leading to a modest cost.
\end{enumerate}
Extensive studies of \CODEXb over a large range of different LLP production and decay mechanisms can be found in the Expression of Interest~\cite{Aielli:2019ivi} and are briefly summarized below.

\begin{figure}[t!]
  \centering
  \includegraphics[width=\linewidth]{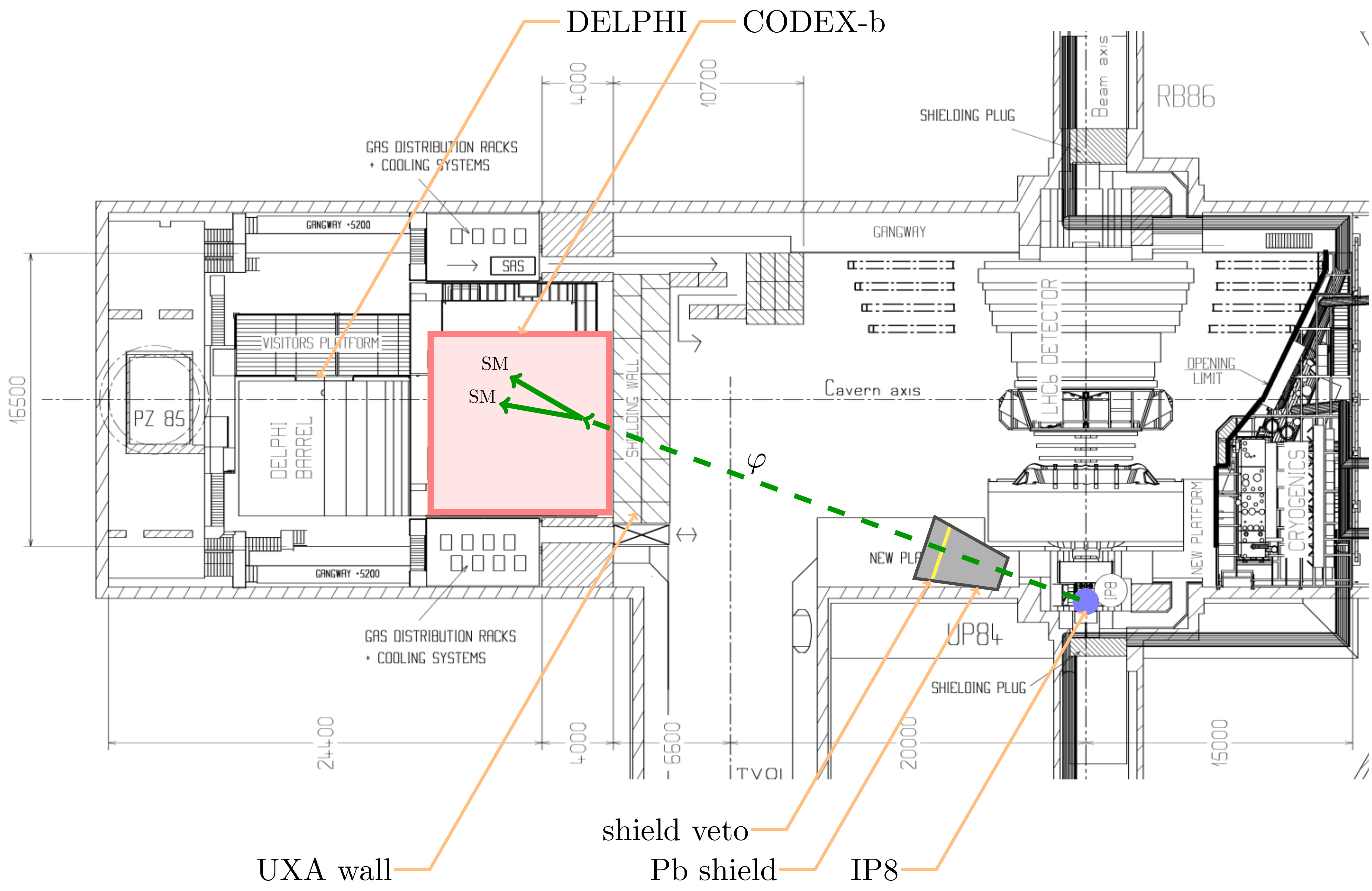}
  \caption{Top view layout of the LHCb experimental cavern UX85 at point 8 of the LHC, overlaid with a schematic of the baseline \CODEXb detector. Adapted from \ccite{Gligorov:2017nwh}.\label{fig:LHCbCav}}
\end{figure}

The proposed \CODEXb detector location is roughly 25 meters from IP8, and the baseline geometry of the detector comprises a fiducial volume of $10\times10\times 10\,\mathrm{m}^3$, as shown in \cref{fig:LHCbCav}.
Backgrounds are controlled by passive shielding provided by a combination of existing passive shielding (the concrete UXA radiation wall) and an array of active vetoes and passive shielding to be installed adjacent to IP8.
A smaller proof-of-concept demonstrator detector, \CODEXbeta, has been installed in the proposed location of \CODEXb, shielded only by the existing wall. 

The remainder of this document is structured as follows. 
We review the scientific context of \CODEXb, including motivation and physics reach in \cref{sec:context}. 
Our methodology towards a successful experiment is described in \cref{sec:methodology}, including simulation and optimization tools and the status of background simulations.
\Cref{sec:readiness} is devoted to the readiness and expected challenges to make \CODEXb a reality,  including a discussion of the features and constraints on realistic detector geometries and their performance.
\Cref{sec:timeline,sec:construction} outline the timeline and costs, respectively. 
Finally, \cref{sec:codexbeta} describes the status of the \CODEXbeta demonstrator detector. 
We conclude in \cref{sec:conclusions}.

\section{Scientific context}\label{sec:context}

Discussed in detail in \ccite{Aielli:2019ivi}, the physics case for \CODEXb is motivated by the very broad class of models that may be probed through LLP searches: almost any model that features a hierarchy of mass scales, loop suppression, and/or small couplings may have an LLP in its spectrum of states.
The broad array of possible LLP signatures, however, raises the problem of how to achieve comprehensive coverage of the theory landscape; this can only be addressed with a set of complementary experiments and searches.
The physics case of \CODEXb may be understood via two complementary approaches, studying: 
\begin{enumerate}
\item \textit{Minimal models}:
  These are simple SM extensions containing a single new particle that is neutral under all SM gauge interactions.
  Such simplified models have limited predictive power and physical interpretation, but are good representatives of more complicated ultraviolet completions that address one or more open problems of the SM. 
  This approach motivated the set of benchmark models developed during the Physics Beyond Colliders (PBC) effort~\cite{Beacham:2019nyx}.
\item \emph{Complete models}:
  These are more complex realizations that address one or more of the open problems in the SM, including, \eg the hierarchy problem, baryogenesis, and dark matter.
\end{enumerate}
In the remainder of this section, we briefly summarize the sensitivities under these two approaches for the \CODEXb baseline design presented in \cref{sec:baseline};
the performance of this design, accounting for tracking and vertex reconstruction efficiencies, is characterized in \cref{sec:readiness} along with the relative performance of more realistic detector configurations.

\subsection{Minimal models}\label{sec:minmodels}

The symmetries of the SM restrict the possible couplings through which a new, neutral state may interact with the SM sector.
A simple classification is thus possible through the spin of the new state. 
One typically considers a scalar ($S$), pseudo-scalar ($a$), a fermion ($N$) or a vector ($A'$), where each allows for a handful of dimension-4 and/or 5 operators:
\begin{subequations}
  \begin{align}
    \text{Abelian hidden sector:\quad}&
    F_{\mu\nu}F'^{\mu\nu},\quad h A'_{\mu}A'^{\mu}\label{eq:vectorportal}\\
    \text{Dark Higgs:\quad}&S^2H^\dagger H, \quad S H^\dagger H\\
    \text{Axion-like particles:\quad}&\partial^\mu a\, \bar \psi
    \gamma_\mu\gamma_5 \psi,\quad a W_{\mu\nu}\tilde W^{\mu\nu},\nonumber\\ 
    & a B_{\mu\nu}\tilde B^{\mu\nu}, \quad a G_{\mu\nu}\tilde G^{\mu\nu}.
    \label{eqn:alpportal} \\
        \text{Heavy neutral leptons:\quad}&\tilde{H} \bar{L}N
    \label{eqn:hnlportal}
  \end{align}
\end{subequations}
Here $F'^{\mu\nu}$ represents the field strength operator to the vector field $A'$;
$H$ the SM Higgs doublet;
$h$ the physical, SM Higgs boson;
$L$ the SM lepton doublets;
$\psi$ any SM fermion;
and $B^{\mu\nu}$, $W^{\mu\nu}$ and $G^{\mu\nu}$ the field strengths of the SM hypercharge, $SU(2)$ and strong forces, respectively. 
We also allow for scenarios where a different operator is responsible for the production and decay of the LLP, as summarized below. Finally, we generally use the definitions and conventions specified in \ccite{Aielli:2019ivi}. 

The \textbf{Abelian hidden sector} model~\cite{Schabinger:2005ei,Gopalakrishna:2008dv,Curtin:2014cca} involves just one additional, massive $U(1)$ gauge boson ($A'$) and its corresponding Higgs boson ($H'$).
The $A'$ and the $H'$ mix with, respectively, the SM photon and the SM Higgs.
If the latter is heavier than the SM Higgs, it decouples from the phenomenology, leaving behind the operators in \cref{eq:vectorportal} in the low-energy effective theory.
Production of the $A'$ occurs through the $h A'_{\mu}A'^{\mu}$ operator, while the $A'$ decay proceeds via kinetic mixing $F_{\mu\nu}F'^{\mu\nu}$.
The production and decay rates of the $A'$ are therefore controlled by independent parameters.
The top left plot of \cref{fig:portals} shows the reach of \CODEXb for a light $A'$ mass.

The \textbf{dark Higgs} or Higgs portal simplified model involves the maximally minimal addition of a single, real scalar degree of freedom ($S$) that couples to the SM Higgs.
The model has three free parameters: the mass ($m_S$), the sine of the mixing angle with the Higgs ($\sin\theta$), and the mixed quartic coupling with the Higgs ($\lambda_D$).
The mixing angle controls the lifetime of $S$ as well as the production rate, which is dominated by exotic $b$-hadron decays.
When non-zero, $\lambda_D$ enables pair production of $S$ both in exotic Higgs and $b$-hadron decays.
LHCb already has sensitivity to this model~\cite{Aaij:2016qsm,Aaij:2015tna}, but \CODEXb would greatly extend the reach into the small-coupling/long lifetime regime, see the bottom row of \cref{fig:portals}.

\begin{figure*}
  \centering
  \includegraphics[width=0.49\textwidth]{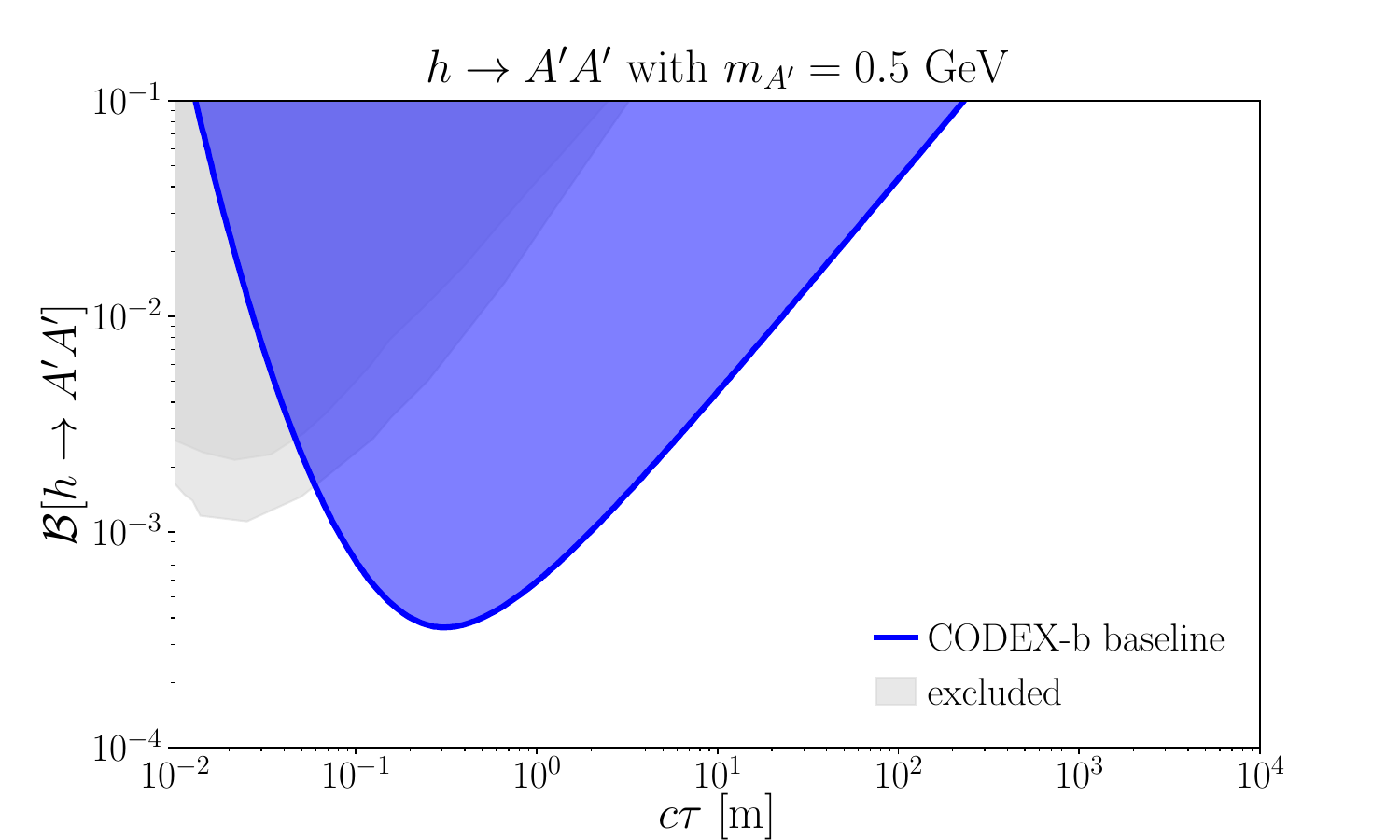}
  \includegraphics[width=0.49\textwidth]{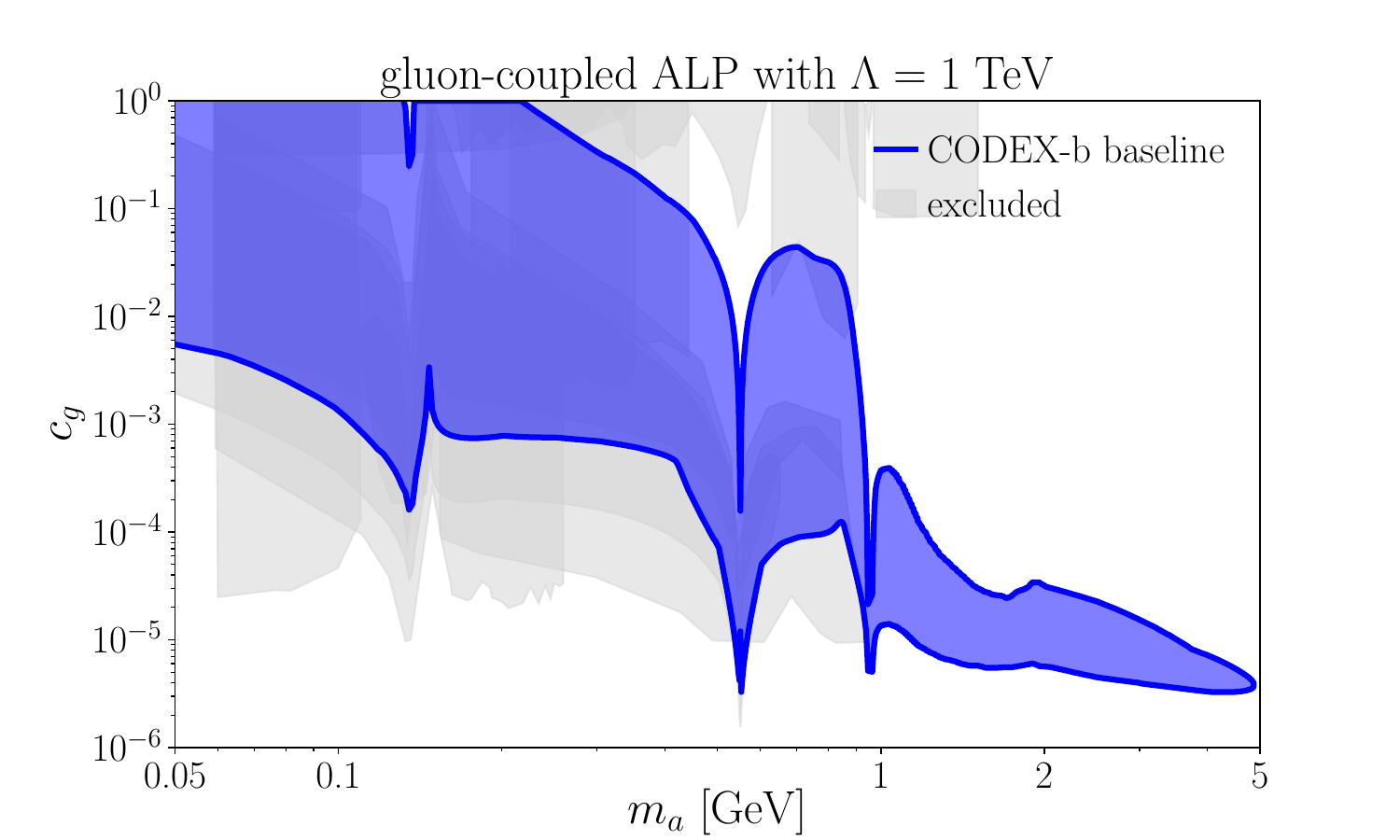}\\
  \includegraphics[width = 0.49\textwidth]{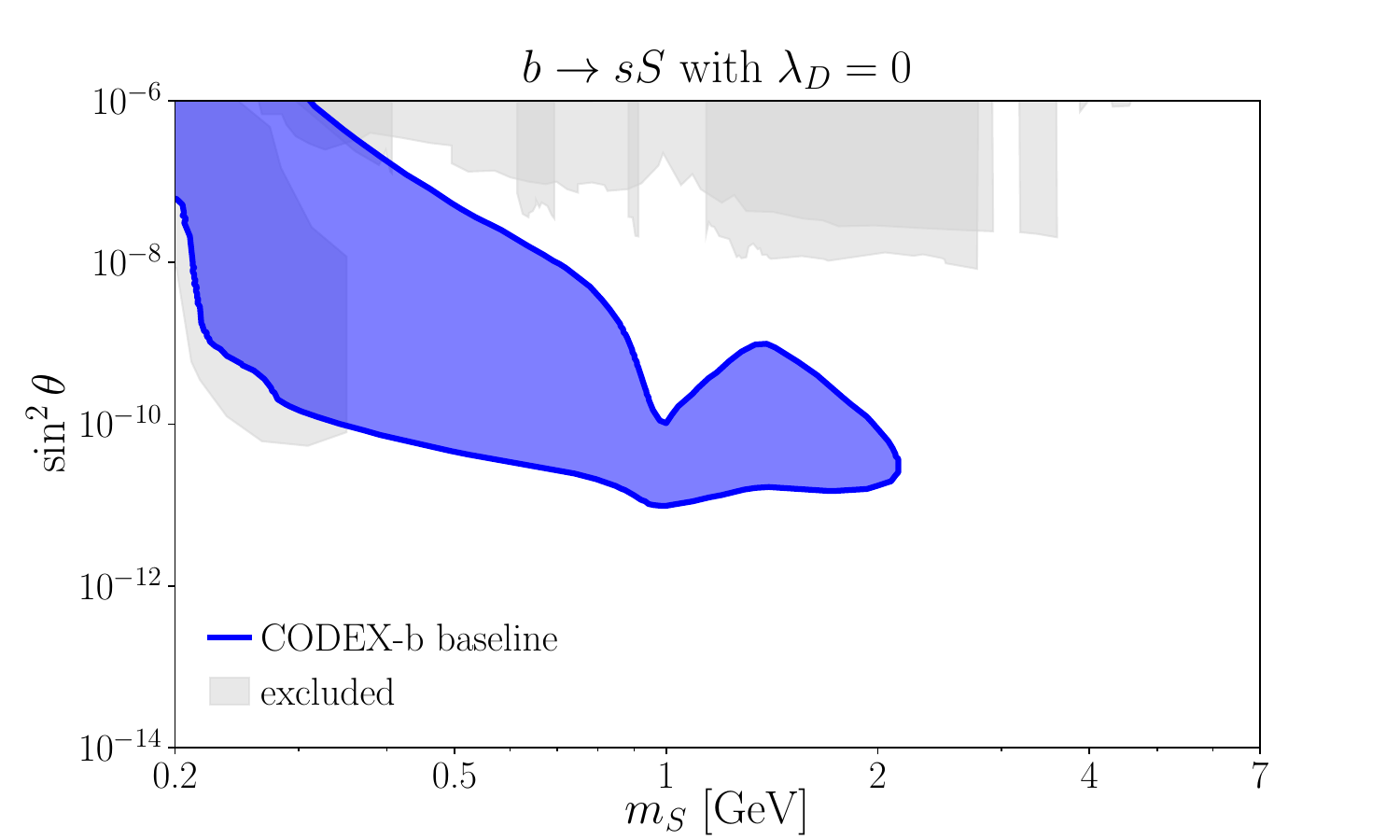}
  \includegraphics[width = 0.49\textwidth]{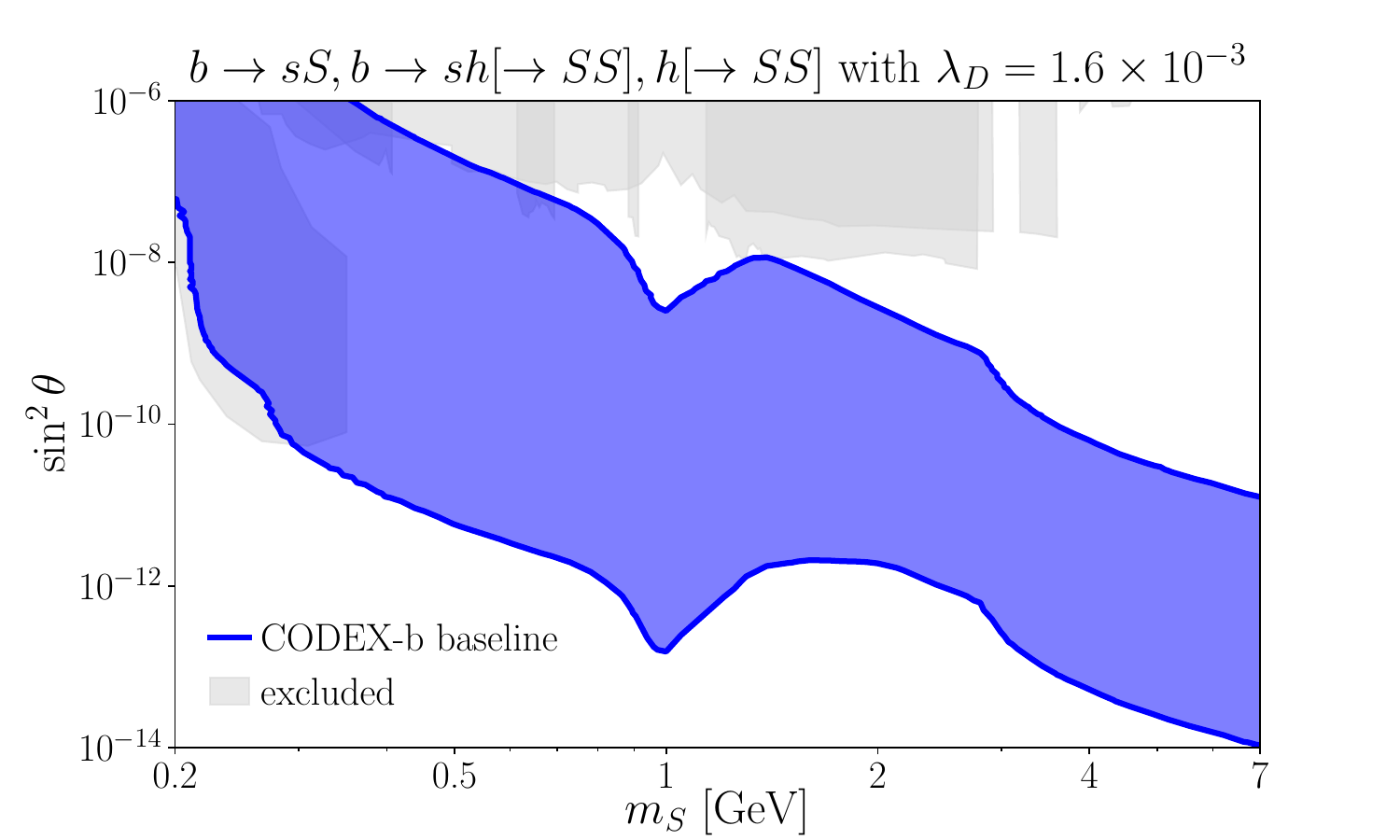}\\
  \caption{\textbf{(top right)} Baseline \CODEXb reach for $h\to A'A'$ . \textbf{(top left)} Gluon-coupled axion-like-particle reach. \textbf{(bottom)} Simplified dark Higgs model reach with the mixed quartic coupling $\lambda_D$ chosen such that (left) $\mathcal{B}[h\to SS]=0$ and (right) $\mathcal{B}[h\to SS]=0.01$. Plots modified from \ccite{Aielli:2019ivi} with updated exclusion limits~\cite{CMS:2021sch,CMS:2024bvl,CMS:2024qxz,FASER:2024bbl}.\label{fig:portals}}
\end{figure*}

\textbf{Axion-like particles} (ALPs) couple to SM fields through dimension-5 operators~\eqref{eqn:alpportal}, which can be generated in a broad range of BSM scenarios.
ALPs tend to be light when generated via the breaking of approximate Peccei-Quinn-like symmetries,  and their suppressed couplings make them natural LLP candidates.
Long-lived ALPs may be abundantly produced at the LHC via their couplings to quarks and/or gluons, through several mechanisms, including: production during hadronization; production from hadron decays via neutral pseudoscalar meson mixing; and production from flavor-changing neutral-current decays of strange or $b$-hadrons.
In addition, for transverse LLP experiments such as \CODEXb, of particular importance for gluon-coupled ALPs is production by emission in the parton shower.
This can lead to substantial enhancements in sensitivity.
The projected ALP sensitivity of \CODEXb for gluon-coupled ALPs is shown in the top-right plot of \cref{fig:portals} and for fermion-coupled ALPs can be found in \ccite{Aielli:2019ivi}.

\textbf{Heavy neutral leptons} (HNLs) couple to the SM via the marginal lepton Yukawa operators in \cref{eqn:hnlportal}, or may arise in many simplified NP models involving higher-dimensional operators.
These may include explanations for the neutrino masses or theories of dark matter, among others (see below).
UV completions of SM-HNL operators imply an active-sterile mixing $\nu_\ell = U_{\ell j} \nu_j + U_{\ell N} N$,  where $N$ is the HNL, and the active-sterile mixing $U_{\ell N}$ is an extended Pontecorvo–Maki–Nakagawa–Sakata (PMNS) neutrino mixing matrix element.
The projected HNL sensitivities of \CODEXb for single-flavor mixing with the $\mu$ and $\tau$ neutrinos can also be found in \ccite{Aielli:2019ivi}.

\subsection{Complete models}

Here we review a few examples of complete models featuring LLPs that aim to explain open problems, such as the hierarchy problem, baryogenesis, or the origin of dark matter, that \CODEXb has the potential to discover. 
For more details, see \ccite{Aielli:2019ivi}.

Models of \textbf{R-parity violating (RPV) supersymmetry} feature light neutralinos, produced through exotic $B$, $D$, or $Z$ decays, that \CODEXb could discover~\cite{Dercks:2018eua,Helo:2018qej}. 
The sensitivity is approximately independent of the flavor structure of the RPV coupling(s), so long as the net branching ratio to at least two charged tracks is not suppressed. 

In \textbf{relaxion models}~\cite{Graham:2015cka}, the light scalar $S$ in the dark Higgs model stabilizes the electroweak scale~\cite{Flacke:2016szy,Choi:2016luu}.

\textbf{Neutral Naturalness} models~\cite{Craig:2014aea,Craig:2015pha} build on the Twin Higgs paradigm~\cite{Chacko:2005pe,Chacko:2005un} that aims to alleviate the SM hierarchy problem. 
Several versions of these models~\cite{Craig:2015pha,Craig:2016kue} produce LLPs in exotic Higgs decays. 

In many dark matter (DM) models, additional, unstable particles beyond the (meta)stable DM itself are needed to achieve the observed DM relic density. 
In many cases, these extra particles may be LLPs, detectable at \CODEXb.
For example, \CODEXb would be sensitive to \textbf{inelastic DM} models~\cite{Tucker-Smith:2001myb, Izaguirre:2015zva} that produce very soft, displaced signatures in collider experiments~\cite{Berlin:2018jbm}. 
DM models with \textbf{sterile co-annihilation}~\cite{DAgnolo:2018wcn} exhibit phenomenology comparable to the dark Higgs simplified benchmark. 
\textbf{Asymmetric DM} models~\cite{Kaplan:2009ag,Kim:2013ivd,Zurek:2013wia} and \textbf{Strongly Interacting Massive Particles (SIMPs)}~\cite{Hochberg:2014dra,Hochberg:2014kqa,Choi:2017zww,Choi:2018iit} implicate the $\gev$ scale, and can contain additional LLP states in the dark sector that could be discovered at \CODEXb. 
In \textbf{Freeze-in models}~\cite{Hall:2009bx}, the DM is never in equilibrium with the SM, which enforces very feeble couplings. 
These models similarly generically predict LLPs that could be accessible to \CODEXb.

Certain models of baryogenesis rely on out-of-equilibrium decays in the early universe and predict LLPs: \textbf{WIMP baryogenesis} is such an example~\cite{Cui:2012jh,Cui:2014twa}.
The baryon asymmetry can also be generated through \textbf{indirect CP-violation} from heavy flavor baryons~\cite{McKeen:2015cuz,Aitken:2017wie}. 
These models predict exotic $b$-hadron decays to LLPs, which may be detectable at \CODEXb. 

Finally, \textbf{HNLs} may arise in explanations for the neutrino masses \cite{Mohapatra:1986bd}, the $\nu$MSM \cite{Asaka:2005an,Asaka:2005pn}, DM models \cite{Batell:2017cmf}, or models which aim to address lepton flavor universality anomalies~\cite{Greljo:2018ogz,Asadi:2018wea,Robinson:2018gza}.

\section{Tools and Methodologies}\label{sec:methodology}

\subsection{Baseline design and principles}\label{sec:baseline}

The baseline detector configuration considered in \ccite{Gligorov:2017nwh,Aielli:2019ivi} comprised sextet resistive plate chamber (RPC) panels on the six outer faces of the $10\times10\times10~\text{m}^{3}$ cubic detector volume, along with four uniformly-spaced internal RPC triplet stations along the $x$ axis (in beamline coordinates).
This cubic volume is located at $x=[26,36]~\text{m}$ (transverse), $y=[-7,3]~\text{m}$ (vertical) and $z = [5,15]~\text{m}$ (forward) with respect to IP8. 
The proposed tracking technology for \CODEXb follows the ATLAS phase-II RPC design~\cite{Collaboration:2285580}, so that tracking stations will be composed of arrays of $1.03\times1.88~\text{m}^2$ triplet RPC modules, supported by a structural aluminum frame.  
The sensitivity curves of \cref{fig:portals} apply to the baseline detector configuration, assuming $100\%$ LLP vertex reconstruction efficiency.
In practice, realistic configurations can achieve vertex reconstruction efficiencies of $40$--$90$\%, which amount to small shifts in the logarithmic parameter spaces shown in \cref{fig:portals}.

In general, an idealized LLP detector design seeks:
\begin{enumerate}[noitemsep]
\item \textbf{Optimal signal sensitivity}: Ensuring the track acceptance is high and instrumentation is capable of reconstructing LLP decay vertices with high efficiency in any given fiducial volume.
\item \textbf{Background suppression}: Requiring passive shielding of primary backgrounds, combined with active vetoes for production of secondary backgrounds in the shielding material (primarily by muons); see \cref{sec:background}.
\end{enumerate}
Based on these drivers, the design principles of the baseline configuration were informed by a very conservative approach, which included: (a) a hermetic design to maximally capture signal tracks, veto incoming muons, and veto potential soft cavern backgrounds and (b) a substantial number of internal tracking stations to ensure hits as close as feasible to the decay vertex and thus high track and vertex resolution.

While the baseline design was shown to permit high vertex reconstruction efficiencies over a broad range of portals~\cite{Gligorov:2017nwh}, it requires a very large amount of instrumented surface: a total of $800$ $1\times2~\text{m}^2$ RPC triplet panels.
The production of the required amount of tracking surface poses a significant challenge, both with respect to the short production timescales and with respect to costs and installation time.
Moreover, as discussed in \cref{sec:readiness}, engineering and technical coordination constraints in the nominal \CODEXb location preempt the possibility of hermetic RPC coverage (though simpler technologies such as scintillator may be possible for the purpose of vetoes).
Thus, relaxing the conservative approach and assumptions for the baseline design is necessary.

\subsection{Fast simulation and optimization frameworks}\label{sec:opt}

To develop more realistic detector design principles and explore the relaxation of the conservative baseline approach, the \CODEXb collaboration has developed a versatile fast simulation framework~\cite{Gorordo:2022rro}.
This enables rapid simulation of the response to variations in the detector geometry and layouts, with respect to various benchmark LLP production and decay channels, across a range of LLP masses and lifetimes.
The studied benchmark scenarios are selected to address a core challenge in optimizing LLP detector design: the broad range of BSM scenarios that may be probed leads to many well-motivated signal morphologies, with widely differing efficiencies and acceptances.
For instance, the LLP boost distribution and decay products vary significantly between the dark-Higgs-portal, and the ALP and Abelian-hidden-sector benchmark models mentioned in \cref{sec:minmodels}.

\begin{figure}
  \centering
  \includegraphics[width=\linewidth]{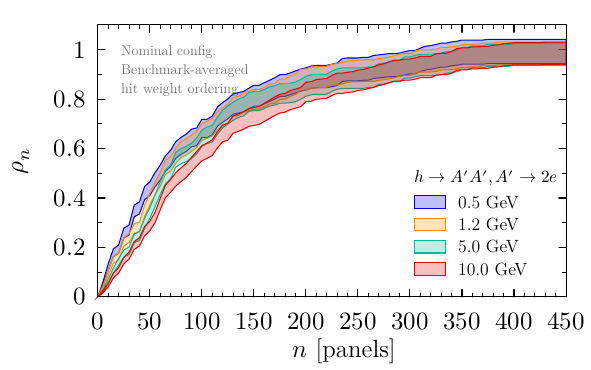}
  \caption{Example relative vertex reconstruction efficiencies ($1\sigma$ CL bands) versus number of panels~\cite{Gorordo:2022rro}. All uncertainties are from the finite statistics of the simulated samples.\label{fig:releffs}}
\end{figure}

In addition, \CODEXb has developed a deterministic optimization framework~\cite{Gorordo:2022rro}  based on a combination of set theoretic and branch-and-bound methods.
This framework is capable of optimizing, in linear time, exponentially large sets of detector configurations in the nominal $10\times10\times10~\text{m}^{3}$ volume subject to realistic engineering or vertex-reconstruction constraints.
Simple hit density estimators are also identified that tend to be excellent estimators for systematically optimized configurations.
As an example, in \cref{fig:releffs} we show the relative vertex reconstruction efficiencies as a function of the number of panels, generated by this estimator, for eleven different benchmarks.
The notable negative curvature for most benchmarks indicates that large reductions in the number of RPC panels, and thus the elimination of hermetic coverage, are typically possible while maintaining good LLP vertex reconstruction efficiencies.

\subsection{Informed design principles and drivers}\label{sec:informed}

The core lesson of these optimization studies is that non-hermetic configurations for \CODEXb exist that can attain good sensitivity over the space of LLP scenarios while reducing the required amount of tracking layers by an $\mathcal{O}(1)$ factor.
This, in turn, permits a significant reduction of the forecasted costs, construction, and installation times for the experiment.

Core drivers for \CODEXb design principles under this informed approach comprise:
\begin{enumerate}[noitemsep]
\item \textbf{Geometric acceptance}: sextet RPC tracking layers are required on the entire $x=36~\text{m}$ back face, but far fewer or no RPCs may be required in practice on the $z=5~\text{m}$ near face, or the $y=3~\text{m}$ top or $y=-7~\text{m}$ bottom faces.
\item \textbf{Vertex resolution}: Good reconstruction resolution of an LLP decay vertex requires at least six hits per track.
For areas where only external triplet rather than sextet layers are possible, adjacent triplet internal layers may be required to ensure acceptable track resolution. 
Internal layers and external sextets are otherwise minimized.
\item \textbf{Backgrounds}: Production of secondary particles in the shielding material (primarily) by muons motivates active veto layers on the front $x=26~\text{m}$ face of the detector.
  Vetoing soft cavern backgrounds may still require hermetic coverage of the detector volume.
  However, because vetoing backgrounds requires good efficiency rather than hit resolution, this may be achieved with cheaper and simpler scintillator technologies, rather than RPCs.
\end{enumerate}

In all simulations of the vertex reconstruction efficiencies for designs informed by these principles, we also require (a) a minimum track threshold of $600~\mev$ because of expected multiple rescattering of softer tracks;
(b) tracking hits must be separated by at least $2~\cm$ on any given layer, corresponding to the finite RPC hit resolution;
(c) an LLP decay vertex resolution of $10~\cm$, \ie at least two reconstructed tracks must pass within $10~\cm$ of each other.

\subsection{Backgrounds and simulation}\label{sec:background}

Collisions at IP8 produce a large flux of background primary hadrons and leptons. 
Of these, primary neutral LLPs, \eg (anti)neutrons and $K_L^0$'s, can enter the detector and decay or scatter into tracks resembling a signal decay.
Suppression of these primary-hadron fluxes can be achieved with passive shielding material: for a shield of thickness $L$, the background flux suppression $\sim e^{-L/\lambda}$ where $\lambda$ is the material nuclear interaction length.
Background flux rate studies have been performed with \textsc{Geant4} \cite{Agostinelli:2002hh} and considering various interaction lengths, reaching an optimal zero-background baseline \CODEXb design~\cite{Aielli:2019ivi} with $3~\text{m}$ of concrete in the UXA radiation wall, corresponding to $7\lambda$ of shielding and supplemented with $4.5~\text{m}$ ($25\lambda$) of Pb shielding.

The large amount of shielding material may act in turn as a source of neutral LLP secondaries, produced mainly by scattering of muons or neutrinos.
The most concerning neutral secondaries are produced $<1~\text{m}$ from the far side of the shield by muons that slow down and stop before reaching the detector itself.
Such muons are invisible to the detector, while their neutral secondaries, such as $K_L^0$'s, may reach the detector volume.

\Ccite{Gligorov:2017nwh,Aielli:2019ivi} have shown that this problem may be solved with an active veto layer in the shield itself, located to sufficiently veto most muons that produce secondaries, but not so close to the IP that all events are vetoed.
Detailed simulations of the setup involve careful treatment of the primary background fluxes at the interaction point, folded into a special \textsc{Geant4} simulation of shielding sub elements --- usually $\sim 1~\text{m}$ thick slices of shielding material --- that encode the propagation and secondary production of dozens of different background particles species, over a large range of energies. 

As for the baseline design, the simulation makes a series of conservative assumptions:
\begin{itemize}[noitemsep]
\item Angular distribution of particle scattering is not exploited;
  all particles scattered within $23^\circ$ of the forward direction, corresponding to the approximate angular acceptance of the detector, are retained.
\item Detector response to neutral secondary particles decaying to charged particles is assumed to be 100\%.
\item Longer path lengths from non-zero angles of incidence on the shield wall are not included.
\item The active veto is implemented in a single layer and does not use tracking information.
\end{itemize}
Relaxing these simulation assumptions would allow for the study of segmented veto layers that are able to exploit directionality to more efficiently veto background fluxes. 
Further, simulation of the detector response to background fluxes can improve understanding of likely background-rejection efficiencies.
Both aspects may be studied with the tools already being developed for the optimization studies in \cref{sec:opt}, to reduce, possibly substantially, the required amount of lead shielding.

As mentioned earlier, a salient feature of the \CODEXb proposal is to trigger on events with an ``interesting'' pattern of hits in both \CODEXb and the main LHCb detector. 
Details of the cavern infrastructure geometry and the LHCb magnetic field must be included in the simulation. 
Backgrounds due to processes other than proton-proton collisions at the LHCb interaction point, known as the LHC machine-induced background (MIB), also occur. 
To accommodate these, a full simulation including LHCb, \CODEXb/\CODEXbeta, cavern infrastructure, and MIB, is being developed.
Details on the full simulation for \CODEXbeta are presented in \cref{ssec:swbeta} and a preliminary setup for Run~1/2 was described in \ccite{Dey:2019vyo}.
This is now being extended to Run~3 data-taking conditions and the \CODEXbeta installation.

The CERN radiation group has placed a battery-operated radiation monitor in the UX85A-D barrack region for Run~3 data taking.
This monitor complements the charged background flux measurements of \ccite{Dey:2019vyo}, and those that will be measured by \CODEXbeta, since the monitor is sensitive to thermal neutrons that are difficult to simulate.

\section{Readiness and expected challenges}\label{sec:readiness}

\subsection{Detector location and representative designs}
 
\begin{figure*}
  \includegraphics[width=0.325\linewidth]{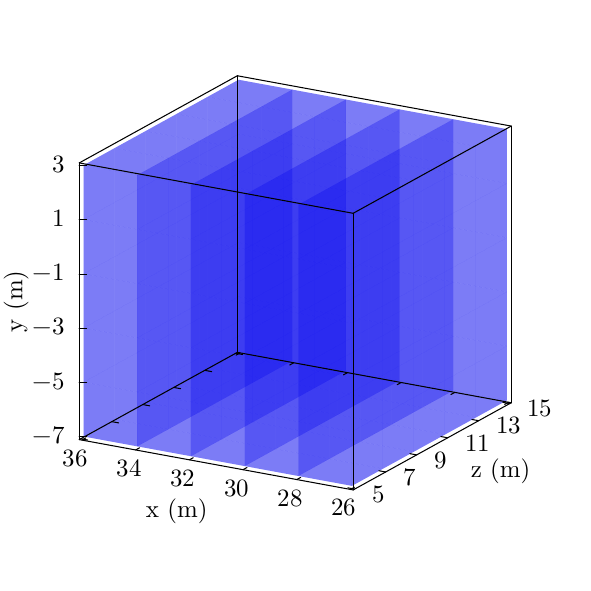}
  \includegraphics[width=0.325\linewidth]{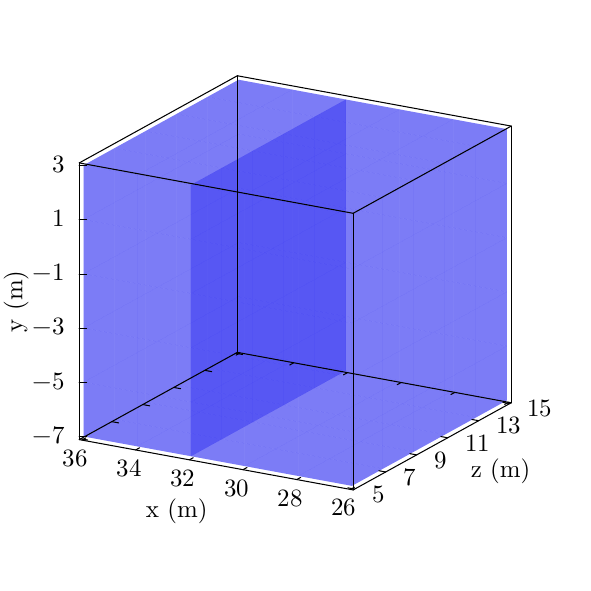}
  \includegraphics[width=0.325\linewidth]{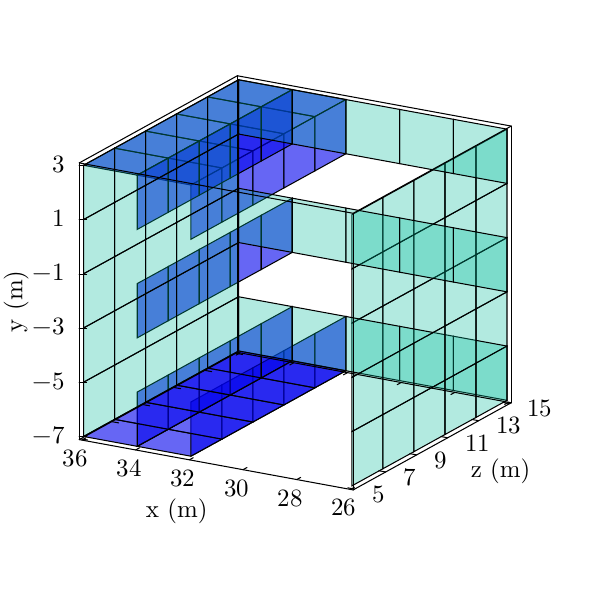}
  \caption{Schematics of \CODEXb design scenarios where triplet (sextet) RPC panels are shown in blue (green):
    \textbf{(left - \cbsB)} baseline configuration with only triplets,
    \textbf{(middle - \cbsC)} baseline configuration with only triplets and a single internal plane,
    \textbf{(right - \cbsD)} an optimized, non-RPC-hermetic configuration subject to cavern constraints.\label{fig:repdesigns}} 
\end{figure*}
 
The proposed detector geometry and location will be primarily informed by the \CODEXbeta demonstrator but is also constrained by engineering and logistical considerations, which are the subject of ongoing deliberations between the LHCb and \CODEXb collaborations. 
The simulation and optimization framework described in \cref{sec:opt} may be used in the development of a realistic design for the full \CODEXb detector, once the engineering constraints in the cavern are fully defined.
The main engineering challenges typically involve the integration of existing architecture in the UXA cavern with the services and mechanical supports required by the detector. 
The difficulty of overcoming these challenges will influence the ultimate selection of location and detector geometry.

To characterize the range of possibilities that may be considered, we show in \cref{fig:repdesigns} three \CODEXb scenarios motivated by the discussion of \cref{sec:informed}.
These scenarios are still under development and serve only to highlight possibilities that might be available as deliberations continue.
\CbsB is a minor variation of the baseline design to remove all sextets in favor of RPC triplet modules.
\CbsC further removes all internal tracking stations except one.
Finally, we consider \cbsD that uses an optimized geometry per \cref{sec:opt,sec:informed} to minimize RPC modules while retaining high sensitivity and vertex resolution.
It represents a more opportunistic scenario, in which the location of the RPCs has been adapted to putative constraints from existing infrastructure in the UXA cavern, thus reducing the need to reconfigure the space significantly.
The \cbsA design requires $800$ RPC triplet modules, \cbsB requires $500$, and \cbsC and \cbsD require $350$.

For \cbsD, the sextets could be replaced with triplets and augmented with scintillator panels for similar performance, requiring a total of $250$ RPC triplet modules.
The \CODEXbeta RPC triplet frame, CX1, could easily be adapted for such a scenario, where the frame is loaded with scintillator and steel, and SiPMs are mounted on the edges through the card slots.

\begin{table}
  \caption{Relative $c\tau$-averaged vertex reconstruction efficiencies compared to the baseline design, for the three representative \CODEXb scenarios and two benchmark models with representative kinematics (see \ccite{Gorordo:2022rro}, the signal kinematics, and thus expected efficiencies, for ALP LLPs interpolate
between those for the Abelian dark sector ``$h \to A' A'$'' and the dark Higgs ``$b \to s S'$'' benchmarks.). Uncertainties are from the finite statistics of the simulated samples.\label{tab:releffs}}
  \setlength{\tabcolsep}{12pt}
  \begin{tabular}{r|l|l|l}
    \toprule
    \multicolumn{1}{c|}{$m_\text{LLP}$} & 
    \multicolumn{3}{c}{scenario} \\
    \multicolumn{1}{c|}{[GeV]} &
    \multicolumn{1}{c|}{1} &
    \multicolumn{1}{c|}{2} &
    \multicolumn{1}{c}{3} \\
    \midrule
    \multicolumn{4}{c}{\textbf{$h \to A' A',  A' \to 2e$}} \\
    \midrule
    0.5  & 0.81(3)  & 0.56(2) & 0.80(3) \\
    1.2  & 0.81(3)  & 0.55(2) & 0.72(3) \\
    5.0  & 0.86(4)  & 0.58(3) & 0.71(3) \\
    10.0 & 0.88(4)  & 0.55(3) & 0.75(4) \\
    \midrule
    \multicolumn{4}{c}{\textbf{$b \to s S',  S' \to 2e$}} \\
    \midrule
    0.5  & 0.94(11) & 0.61(8) & 0.77(9) \\
    1.0  & 0.94(11) & 0.55(7) & 0.74(9) \\
    2.5  & 0.85(10) & 0.33(5) & 0.53(7) \\
    4.0  & 0.81(11) & 0.22(4) & 0.42(6) \\
    \bottomrule
  \end{tabular}
\end{table}

In \cref{tab:releffs}, we show the $c\tau$-averaged vertex reconstruction efficiencies of each of these variations \textit{relative} to the \cbsA design performance for two representative kinematic scenarios covering the models of \cref{fig:portals};
the kinematics of the ALP portal are roughly an interpolation between the kinematics of the dark photon and dark Higgs sectors.
The vertex reconstruction efficiencies are determined according to the simulation requirements in \cref{sec:informed}.
The studied configurations can achieve 50--90\% relative vertex reconstruction efficiency with respect to the baseline, even with half the instrumentation.
Consequently, a full detector may be realized in the nominal detector location, 
subject to realistic engineering, cost, time, and other technical constraints,  but with performance comparable to the original baseline proposal.
Correspondingly, the baseline sensitivity curves shown in this document (and in \ccite{Gligorov:2017nwh,Aielli:2019ivi}) serve as plausible approximate upper limits on the sensitivity of the full detector and are not expected to shift significantly.

\subsection{Detector technologies}
 
While the RPC design itself is well-established, it is necessary to develop firmware and software to communicate seamlessly between the \CODEXb RPCs, which follow the ATLAS phase-II design, and the LHCb software-only trigger.
This is also necessary for the successful deployment of the \CODEXbeta demonstrator and is already underway with an expected completion before the end of 2025; see \cref{sec:codexbeta}.

The default gas mixture for the \CODEXb RPCs includes R-134a and SF6~\cite{CODEX-b:2024tdl}, both of which have high Global Warming Potential (GWP).
However, ECO65 gas provides a performant alternative~\cite{Proto:2024wjz}.
Given the smaller size of \CODEXbeta~\cite{CODEX-b:2024tdl} and its gas recirculation system, both the default mixture and ECO65 can be tested with minimal environmental impact.
Using these tests to establish the RPC performance, it is expected that \CODEXb will also utilize a full recirculation system and an eco-gas mixture.

\section{Timeline}\label{sec:timeline}

The absolute timeline for \CODEXb depends upon CERN approval and the securing of funding. Consequently, a relative timeline is given here based on our experience building \CODEXbeta.
The \CODEXb proposal does not require any research and development beyond that for \CODEXbeta, except the near-detector shield and an informed design of the support structure.
This shield will consist of lead, with an active scintillator veto, within a movable mechanical support.
Members of the \CODEXb collaboration have expertise in scintillator vetoes through the HeRSCheL detector for LHCb~\cite{Akiba:2018neu}.
The collaboration has also demonstrated mechanical expertise from \CODEXbeta with the development of the \CODEXb module frames, transport carts, and support structure.
These mechanics required tolerances within millimeters and were developed in collaboration with civil engineers.

The two primary time-limiting factors are the production of the RPC modules and the installation of the modules in the physical space.
With a team of $3$ trained personnel, building a single module of $3$ RPCs takes approximately $1$ week.
For \CODEXbeta, our personnel teams consisted of undergraduate students (engineers and physicists), early PhD students (physicists), and post-doctoral researchers (physicists).
For \CODEXb, we will recruit effort from a similarly diverse group of personnel.
The installation was performed with approximately $5$ full-time-equivalent (FTE) for $2$ months.

\begin{table}
  \centering
  \caption{Personnel time estimate for the \CODEXb design scenarios, separated into module building, installation, shield construction, integration with LHCb, and software development. A $20\%$ contingency is included.\label{tab:personnel}}
  \begin{tabular}{l|r|r|r}
    \toprule
    \multicolumn{1}{c|}{task} & \multicolumn{3}{c}{time [FTE years]} \\
    & \cbsA & \cbsB & scenario 2/3 \\
    \midrule
    RPC modules       & 55  & 34 & 24 \\
    detector install  & 52  & 32 & 23 \\
    shield            & 4   & 4  & 4  \\
    \midrule
    software          & 12  & 12 & 12 \\
    LHCb integration  & 11  & 11 & 11 \\  
    \midrule
    total             & 134 & 93 & 74 \\
    \bottomrule
  \end{tabular}
\end{table}

In \cref{tab:personnel}, an estimate is given for the necessary FTE years ($260$ days) to construct and install \CODEXb for the three different scenarios of \cref{fig:repdesigns}, separated into three large-scale tasks: module building, installation, and shielding.
The module building and installation estimates are linearly scaled from \CODEXbeta, and a $20\%$ contingency is included for all estimates.
Most of this FTE, roughly $90\%$, can be provided by student effort as demonstrated with \CODEXbeta.
Frame machining is not included in this estimate, but given its straightforward design, frame production would be best achieved via an external vendor.
We also expect the RPC module building is an overestimate.
Already, the \CODEXbeta team has taken steps towards scaling RPC production.

All three tasks can be performed in parallel.
Currently, the \CODEXb collaboration is $22$ participating institutes and growing.
If each institute provides $3$ FTE years over $2$ years, then all three scenarios except the baseline can be achieved within this period.
If we target the start of $2030$ for a full install, then personnel training and material requisitioning need to begin mid-$2027$.
A staged installation could also be performed, as the installation, given appropriate safety measures, could continue during LHC operation.

\section{Construction and operational costs}\label{sec:construction}

In \cref{tab:costs}, we summarize the construction and operational costs for \CODEXb, including a $20\%$ contingency. The estimated personnel requirements are described in \cref{sec:timeline}.

\begin{table}
  \centering
  \caption{Construction and operational cost estimate for the \CODEXb design scenarios, including a $20\%$ contingency. Numbers in brackets are the standard gas mixture costs (not ECO65).\label{tab:costs}}
  \begin{tabular}{l|r|r|r}
    \toprule
    \multicolumn{1}{c|}{component}
    & \multicolumn{3}{c}{cost [EUR$\times10^6$]} \\
    & \cbsA & \cbsB & scenario 2/3 \\
    \midrule
    RPC modules         & 14.4 &  9.0 &  6.3 \\
    power system        &  3.4 &  2.1 &  1.5 \\
    gas system          &  2.7 &  1.7 &  1.2 \\
    support structure   &  1.4 &  0.9 &  0.6 \\
    DAQ                 &  0.7 &  0.4 &  0.3 \\
    shielding           &  1.2 &  1.2 &  1.2 \\
    tooling             &  1.0 &  1.0 &  1.0 \\
    \midrule
    gas                 & 26.4 & 16.5 & 11.6 \\
    (non-eco gas)       &  3.8 &  2.4 &  1.6 \\
    \midrule
    total               & 51.2 & 32.8 & 23.6 \\
    (non-eco gas)       & 28.6 & 18.7 & 13.8 \\
    \bottomrule
  \end{tabular}
\end{table}

Given our experience developed in building \CODEXbeta, we have a well-defined cost per RPC module of $1.5\times10^4$ EUR based on material purchases.
This is the full cost of the module with the component costs detailed in \cref{tab:bis7}.
The cost of the power system, high and low voltage, is also similarly well defined from \CODEXbeta.

\begin{table}
  \centering
  \caption{Cost of a BIS7-S module.\label{tab:bis7}}
  \begin{tabular}{l|r|r|r}
    \toprule
    item & units & unit cost & total cost [EUR] \\
    \midrule
    $\eta$-panel & 3 & 500 & 1500 \\
    $\phi$-panel & 3 & 500 & 1500 \\
    FE board & 36 & 83 & 3000 \\
    gas gaps & 3 & 1000 & 3000 \\
    frame & 1 & 600 & 600 \\
    cabling & - & - & 400 \\
    DCT & 1 & 4000 & 4000 \\ 
    \midrule
    total & & & 15000 \\
    \bottomrule
  \end{tabular}
\end{table}

For both the gas system and support structure, we have assumed that the cost will scale linearly with the number of RPC modules from the \CODEXbeta cost; this should be an overestimate.
The numbers here assume an aluminum support structure, but steel will need to be used in places;
this will reduce the cost of the support structure.
The DAQ cost is less well defined, given that the future cost of the PCIe40 card (see \ccite{CODEX-b:2024tdl}) is unknown.
However, we assume the cost of each event-builder server is roughly $5\times10^3$ EUR and each PCIe40 is also $5\times10^3$ EUR, although the latter is almost certainly an overestimate.
The cost of the shield is conservatively estimated from the expected cost of the required mechanical structure, lead, and scintillating detector.

The final cost is for the necessary gas over the full lifetime of the detector, assuming a recirculating gas system.
This is the primary operating cost, and is also the largest cost of the experiment other than the RPC modules.
The gas cost estimate here is scaled from \CODEXbeta usage and assumes the EC065 gas mixture.
If running with the $30\%$ CO2 RPC gas mixture, the gas operational cost is expected to decrease by a factor of roughly 7.
Given the importance of the eco-gas mixtures for ATLAS operation, it is expected that the final eco-gas mixture will have a cost significant reduction compared to the current ECO65 mixture.

\section{CODEX-$\beta$}\label{sec:codexbeta}

\subsection{Goals}

\begin{table}
  \centering
  \caption{Total neutral and $K^0_S$ multitrack production during Run~3 in the \CODEXbeta volume for $\mathcal{L} = 15~\text{fb}^{-1}$, requiring $E_{\text{kin}} > 0.4~\gev$ per track~\cite{Aielli:2019ivi}.\label{tab:bkg-tracks}}  
  \setlength{\tabcolsep}{7pt}
  \begin{tabular}{r|r|r}
    \toprule
    \multicolumn{1}{c|}{tracks} 
    & \multicolumn{1}{c|}{total} 
    & \multicolumn{1}{c}{$K^0_L~\text{contribution}$} \\
    \midrule
    1  & $(3.87 \pm 0.11) \times 10^{8}$ & $(2.94 \pm 0.07) \times 10^{8}$ \\
    2  & $(4.09 \pm 0.13) \times 10^{7}$ & $(3.74 \pm 0.13) \times 10^{7}$ \\
    3  & $(5.96 \pm 1.01) \times 10^{5}$ & $(2.92 \pm 0.45) \times 10^{5}$ \\
    4+ & $(9.34 \pm 2.10) \times 10^{4}$ & $(7.03 \pm 1.91) \times 10^{4}$ \\
    \bottomrule
  \end{tabular}
\end{table}

The \CODEXbeta detector, described in \ccite{CODEX-b:2024tdl}, is a small-scale demonstrator for the full-scale \CODEXb detector. 
The primary design goal of \CODEXbeta is to validate the key concepts that justify the building and operation of \CODEXb.
Specifically:
\begin{enumerate}[noitemsep]
\item  Demonstrate the suitability of the mechanical support required for the RPCs and its scalability to the full \CODEXb detector.
\item  Validate the preliminary background estimates from the \CODEXb proposal and the $2018$ background measurement campaign~\cite{Dey:2019vyo}, 
including measurement of soft cavern backgrounds, demonstrating that \CODEXb can be operated as a zero-background experiment (see \cref{sec:background}). 
\item Demonstrate the seamless integration of the detector with the LHCb readout, so that candidate LLP events in \CODEXb can be tagged with the corresponding LHCb detector information to aid in their interpretation. 
\item Demonstrate the suitability of RPCs as a baseline tracking technology for \CODEXb in terms of spatial granularity, hermeticity, and timing resolution.
\item  Demonstrate an ability to reconstruct known SM backgrounds expected to decay inside UX-85A (the proposed location for \CODEXb and \CODEXbeta) and validate a full simulation of the LHCb detector and cavern environment with these measured backgrounds.
\end{enumerate}
The demonstrator built our expertise in building and operating RPCs and the development of the data acquisition, simulation, and software needed to operate \CODEXbeta will significantly simplify similar activities needed for \CODEXb.

A powerful tool to calibrate the detector reconstruction and the RPC timing resolution is the measurement of long-lived SM particles decaying inside the detector acceptance.
The most natural candidates are $K^0_S$ mesons.  
\Cref{tab:bkg-tracks} summarizes the expected multitrack production from decay or scattering on air-by-neutral fluxes entering \CODEXbeta, as well as the contribution from just $K^0_S$ mesons entering the detector. 
Approximately a $\text{few} \times 10^7$ $ K^0_S$ decays to two or more tracks are expected in the \CODEXbeta volume per nominal year of data taking in Run~3,  so that \CODEXbeta will have the opportunity to reconstruct a variety of $K^0_S$ decays.  
The measurement of the decay vertex and decay product trajectories will allow the boost of the LLP to be reconstructed independently from the time-of-flight information.  
Moreover, measurements of the distribution of $K^0_S$ decay vertices can be compared to expectations from the background simulation of the expected $K^0_S$ boost distribution folded against the $K^0_S$ lifetime, allowing calibration and validation of our detector simulation and reconstruction.
Conversely, one may combine the predicted boost distribution and measured vertex distribution to extract the $K^0_S$ lifetime itself.

\begin{figure}
  \centering
  \includegraphics[width=\linewidth]{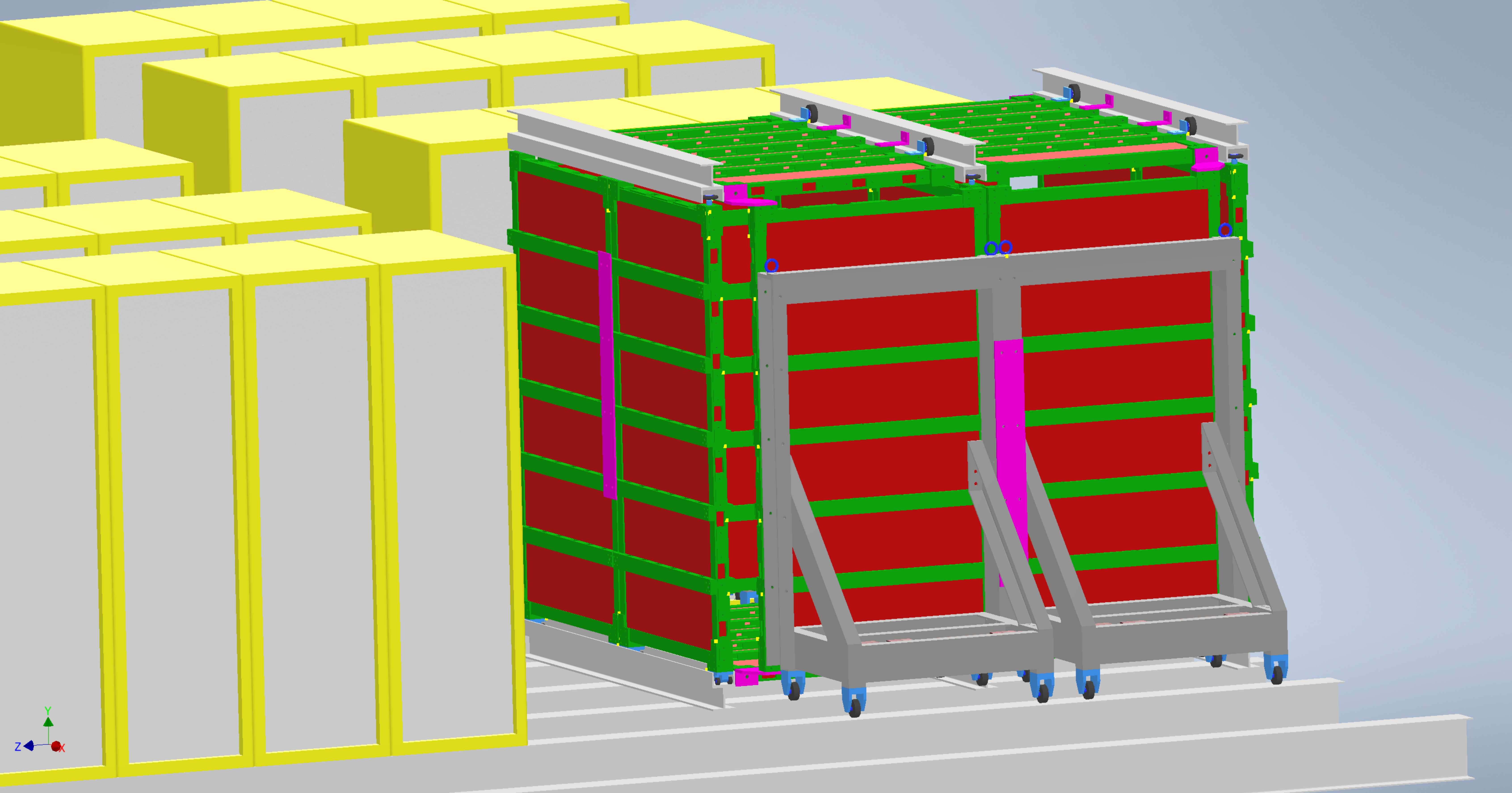}
  \caption{Schematic of \CODEXbeta (green, red, gray) located between the server racks (yellow, gray).\label{fig:demosketch}}
\end{figure}

\subsection{Design and status}

\begin{figure}
  \centering
  \includegraphics[width=\linewidth]{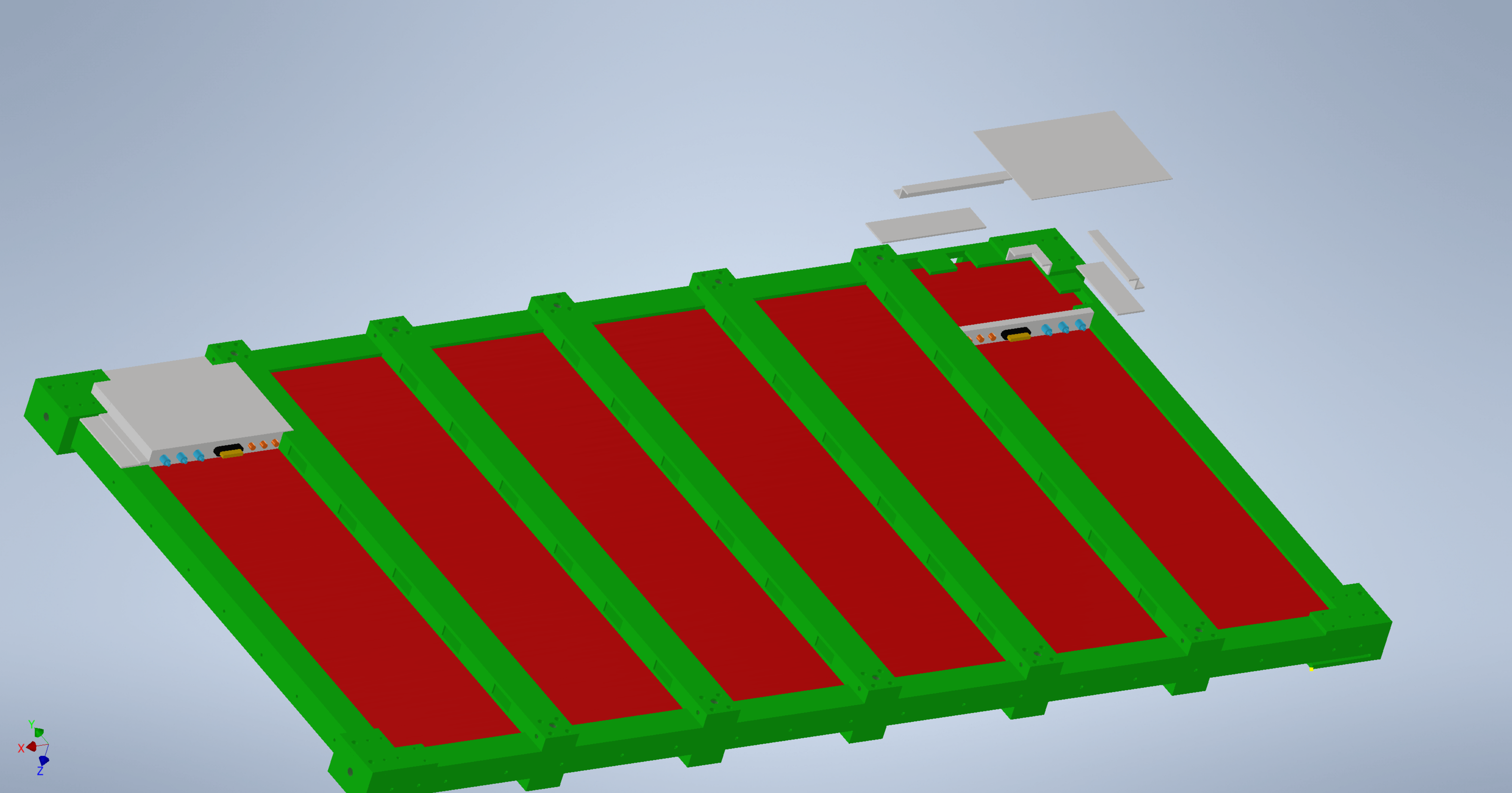}
  \caption{Schematic of a \CODEXbeta module, including the RPC triplet (red), support frame (green), and service boxes (gray).\label{fig:rpc}}
\end{figure}

As shown in the schematic of \cref{fig:demosketch}, \CODEXbeta comprises a $2\times2\times2~\text{m}^3$ cube with an additional interior face.   
Each face contains two modules, each housing a stacked triplet of $2\times1~\text{m}^2$ RPCs integrated into a self-contained mechanical frame.
A total of $(6 + 1)\times2\times3 = 42$ RPC singlets are integrated into a total of $14$ modules.
A schematic of the mechanical frame, specifically designed to withstand the stresses of handling and mounting, is shown in the top diagram of \cref{fig:rpc}.

The location for \CODEXbeta, as well as planned \CODEXb locations, is challenging to access.
Therefore, the frame has been designed to be highly modular and can be assembled with just fastening hardware and no welding.
One of the key installation features is rolling support carts, which can be used to move the modules to the frame.
The final four modules remain in carts, allowing access to the interior of the detector.

The base element of a module is the singlet RPC, identical to the BIS7 RPCs used in the ATLAS Upgrade experiment~\cite{Massa:2716316}.
Each module contains three independent singlets, a triplet, which provide a three-point track and operate in a self-trigger mode, obviating the need for any external reference system for cosmic muon detection.

To form a full $2\times 2\text{m}^2$ face of the \CODEXbeta cube, two modules are placed side-by-side along their $\phi$ sides such that the readouts are on the opposite outer edges of the module.
This maintains the hermeticity of the detector.
For the full \CODEXb design, internal routing of the readout cables may be modified for the sequential positioning of more than two modules.

\begin{figure}
  \centering
  \includegraphics[width=\linewidth]{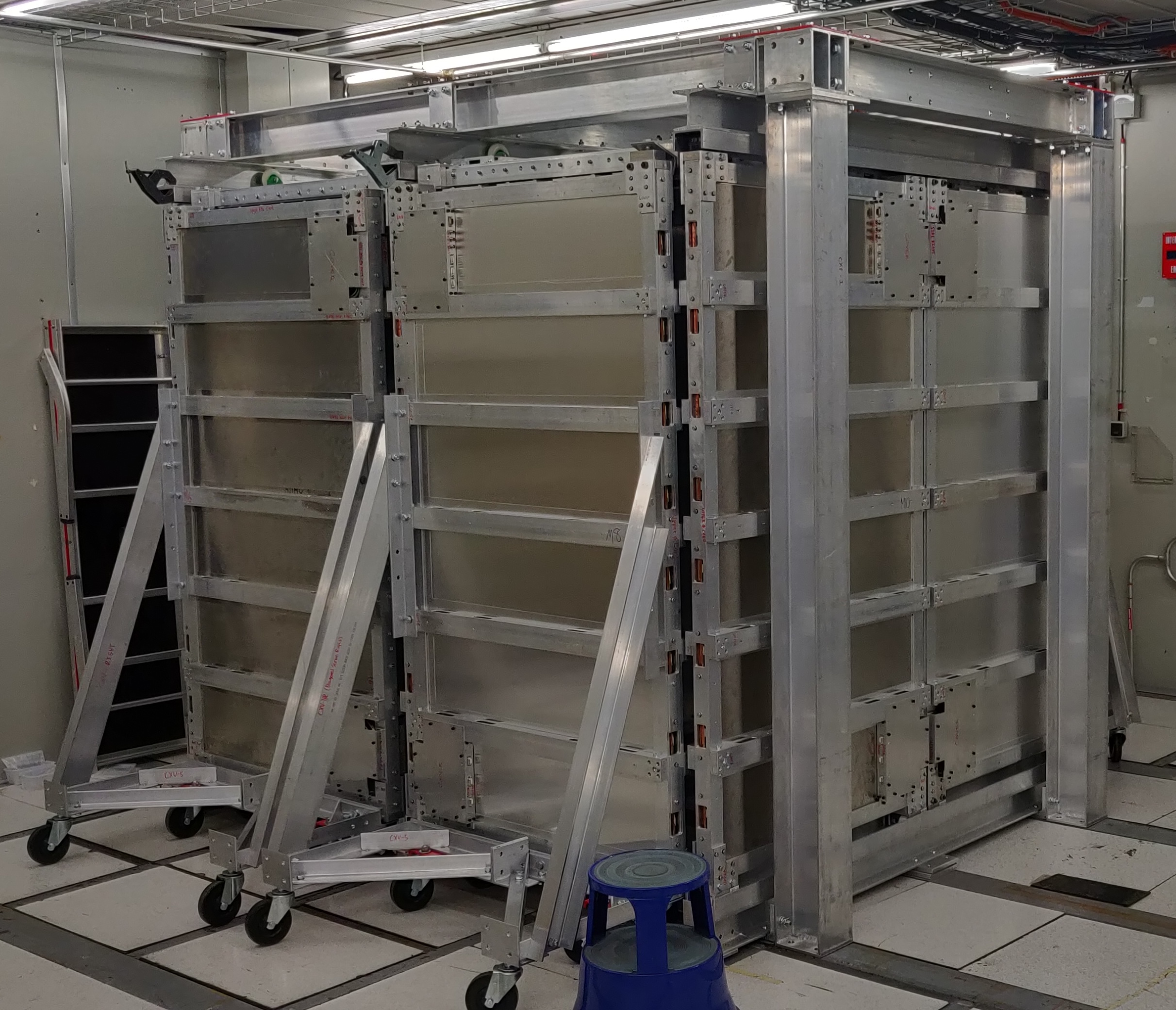}
  \caption{Physical installation of \CODEXbeta at IP8.\label{fig:install}}
\end{figure}

At the time of submission for this document, the \CODEXbeta detector is installed in its location, see \cref{fig:install}.
The power and gas services are prepared, and a test data acquisition system is installed to verify the efficiencies and the noise level of each RPC \textit{in situ}.
All components for the final data acquisition have been designed and are in the process of being procured.
The realization of the full reconstruction and the integration into the LHCb readout require dedicated firmware modifications to the front-end and the LHCb data acquisition.
These developments, together with the necessary software amendments in the LHCb framework to allow simultaneous readout, are in progress.

With the beginning of the data taking in $2025$, we will first commission the hardware with a test data acquisition, validating the performance of the RPCs and measuring background rates with the chambers \textit{in situ}.
Once the final data acquisition system is ready, reconstruction and integration milestones can be verified.

\subsection{Software development and LHCb integration}\label{ssec:swbeta}

A dedicated full software framework for \CODEXbeta is now under preparation. This framework covers the following:

\begin{enumerate}[noitemsep]
\item \textbf{\CODEXbeta event definition}: New data banks and encoders to accommodate the response from the RPC readout into the LHCb DAQ will be developed.
\item \textbf{Simulation}: New Pythia settings, as well as both the \CODEXbeta geometry and a parameterization of the existing concrete wall, will be implemented to simulate any material interaction.
\item \textbf{Digitization}: All tracks from simulation should be digitized to emulate the RPC readout, matching the response format from real data acquisition.
\item \textbf{Reconstruction}: A decoder and algorithms for track reconstruction will be developed, storing data into a suitable format for data analysis.
\item \textbf{Data analysis}: Dedicated analysis tools will be developed for offline data analysis and online data quality assessment during data-taking, in an automated way.
\end{enumerate}
Items 1 to 3 will be implemented into the LHCb software stack, while 4 and 5 use the output from the previous steps in a standalone manner, decoupled from LHCb software.
The complete software framework will be tested and validated before integration into the LHCb software stack.

Before the steps above, the \CODEXbeta readout will be integrated into the LHCb data stream, and we will validate the data acquisition model and its scalability to the full detector.
Currently, the \CODEXbeta event definition and digitization are being developed, and should be completed by the end of summer.
The integration of \CODEXbeta within the LHCb simulation framework is nearing completion, while the event reconstruction will be performed in parallel with other tasks this summer.
Finally, data analysis will be performed during data taking.

\section{Conclusions}\label{sec:conclusions}

Currently, numerous experimental collaborations are working to develop dedicated facilities for studying LLPs, capitalizing on the LHC's production capabilities and luminosity projected for the next twenty years.
Within this research landscape, the \CODEXb experiment occupies a distinctive position: it offers a relatively modest-sized, cost-effective solution using established technologies.
Its strategic location minimizes staffing requirements while enabling integration with LHCb's existing data acquisition systems.
The detector provides competitive or complementary sensitivity for detecting LLPs produced in both transverse and longitudinal directions.
Having confirmed most of the fundamental assertions from its 2017 proposal~\cite{Gligorov:2017nwh}, the \CODEXb collaboration is nearly ready to start taking data with the \CODEXbeta demonstrator --- laying groundwork for the complete \CODEXb detector's construction and implementation in the late 2020s.

Our current focus centers on commissioning the \CODEXbeta demonstrator to capture a significant portion of Run~3 luminosity at IP8.
This collected data will enable precise measurements of both overall background levels in UX-85A and their spatial distribution, informing the optimization of size and configuration for \CODEXb's required active veto shield.
Reconstructing SM backgrounds will provide crucial validation for \CODEXb simulations, allowing confident optimization of the spatial and temporal resolution requirements for \CODEXb RPCs, as well as their physical arrangement.

Following completion of \CODEXbeta activities, we will finalize a fully optimized configuration for both \CODEXb and its active shield veto, comprehensive performance specifications for the \CODEXb RPCs, and a detailed mechanical design for the final experiment.
Based on these developments, we will create a Technical Design Report and pursue a phased construction and installation approach, leveraging UX-85A's shielded environment that permits work during both operational periods and scheduled LHC maintenance shutdowns.

The HL-LHC holds great potential for the exploration of BSM scenarios involving LLPs.
The \CODEXb detector is well-positioned to realize this potential.

\clearpage
\onecolumngrid

\section{Acknowledgments}
We are grateful for the support we have received from the LHCb collaboration. 
We thank the technical and administrative staffs at CERN and at other CODEX-b institutes
for their contributions to the success of the CODEX-b and CODEX-$\beta$ effort.
In addition, we gratefully acknowledge the computing centers and personnel of the
Worldwide LHC Computing Grid and other centres for delivering so effectively the computing infrastructure essential to our analyses.

Individuals have received support from the Science and Technology Facilities Council (UK) [grant reference ST/W004305/1]
and National Research, Development and Innovation Office (NKFIH) research grant (Hungary) [contract number TKP2021-NKTA-64].
The work of IGFAE members and of CVS is supported by the Spanish Research State Agency under projects PID2022-139514NA-C33 and PCI2023-145984-2. The work of IGFAE members is supported 
by Xunta de Galicia (Centro singular de investigación de Galicia accreditation 2019-2022), by European Union ERDF;
and by the Spanish Ministry of Universities under the NextGenerationEU program.
The work of LBNL staff/researchers is supported by the Office of High Energy Physics of the U.S. Department of Energy under contract DE-AC02-05CH11231.
The work of the University of Cincinnati personnel is in part supported by the United States National Science Foundation under grant NSF-PHY-220976.
The work of the DESY personnel is supported by DESY (Hamburg, Germany), a member of the Helmholtz Association HGF, and by the Deutsche Forschungsgemeinschaft
(DFG, German Research Foundation) under Germany’s Excellence Strategy -- EXC 2121 ``Quantum Universe'' -- 390833306.
The work of CVS is supported by Agencia Estatal de Investigación (Spain) through the Ramón y Cajal program RYC2023-043804-I and the Universidade da Coruña-InTalent program.

\renewcommand{\addcontentsline}[3]{}
\section*{References}
\bibliographystyle{JHEP}
\bibliography{main}

\providecommand{\href}[2]{#2}\begingroup\raggedright\begin{thebibliography}{10}

\bibitem{Alimena:2019zri}
J.~Alimena et~al., \emph{{Searching for long-lived particles beyond the Standard Model at the Large Hadron Collider}}, \href{http://dx.doi.org/10.1088/1361-6471/ab4574}{\emph{J. Phys. G} {\bfseries 47} (2020) 090501}, [\href{https://arxiv.org/abs/1903.04497}{{\ttfamily 1903.04497}}].

\bibitem{CMS:2021sch}
{\scshape CMS} collaboration, A.~Tumasyan et~al., \emph{{Search for long-lived particles decaying into muon pairs in proton-proton collisions at $ \sqrt{s} $ = 13 TeV collected with a dedicated high-rate data stream}}, \href{http://dx.doi.org/10.1007/JHEP04(2022)062}{\emph{JHEP} {\bfseries 04} (2022) 062}, [\href{https://arxiv.org/abs/2112.13769}{{\ttfamily 2112.13769}}].

\bibitem{CMS:2021juv}
{\scshape CMS} collaboration, A.~Tumasyan et~al., \emph{{Search for Long-Lived Particles Decaying in the CMS End Cap Muon Detectors in Proton-Proton Collisions at $\sqrt s$ =13\,\,TeV}}, \href{http://dx.doi.org/10.1103/PhysRevLett.127.261804}{\emph{Phys. Rev. Lett.} {\bfseries 127} (2021) 261804}, [\href{https://arxiv.org/abs/2107.04838}{{\ttfamily 2107.04838}}].

\bibitem{Gligorov:2017nwh}
V.~V. Gligorov, S.~Knapen, M.~Papucci and D.~J. Robinson, \emph{{Searching for Long-lived Particles: A Compact Detector for Exotics at LHCb}}, \href{http://dx.doi.org/10.1103/PhysRevD.97.015023}{\emph{Phys. Rev.} {\bfseries D97} (2018) 015023}, [\href{https://arxiv.org/abs/1708.09395}{{\ttfamily 1708.09395}}].

\bibitem{Aielli:2019ivi}
G.~Aielli et~al., \emph{{Expression of interest for the CODEX-b detector}}, \href{http://dx.doi.org/10.1140/epjc/s10052-020-08711-3}{\emph{Eur. Phys. J. C} {\bfseries 80} (2020) 1177}, [\href{https://arxiv.org/abs/1911.00481}{{\ttfamily 1911.00481}}].

\bibitem{Dey:2019vyo}
B.~Dey, J.~Lee, V.~Coco and C.-S. Moon, \emph{{Background studies for the CODEX-b experiment: measurements and simulation}},  \href{https://arxiv.org/abs/1912.03846}{{\ttfamily 1912.03846}}.

\bibitem{Beacham:2019nyx}
J.~Beacham et~al., \emph{{Physics Beyond Colliders at CERN: Beyond the Standard Model Working Group Report}}, \href{http://dx.doi.org/10.1088/1361-6471/ab4cd2}{\emph{J. Phys. G} {\bfseries 47} (2020) 010501}, [\href{https://arxiv.org/abs/1901.09966}{{\ttfamily 1901.09966}}].

\bibitem{Schabinger:2005ei}
R.~M. Schabinger and J.~D. Wells, \emph{{A Minimal spontaneously broken hidden sector and its impact on Higgs boson physics at the large hadron collider}}, \href{http://dx.doi.org/10.1103/PhysRevD.72.093007}{\emph{Phys. Rev.} {\bfseries D72} (2005) 093007}, [\href{https://arxiv.org/abs/hep-ph/0509209}{{\ttfamily hep-ph/0509209}}].

\bibitem{Gopalakrishna:2008dv}
S.~Gopalakrishna, S.~Jung and J.~D. Wells, \emph{{Higgs boson decays to four fermions through an abelian hidden sector}}, \href{http://dx.doi.org/10.1103/PhysRevD.78.055002}{\emph{Phys. Rev.} {\bfseries D78} (2008) 055002}, [\href{https://arxiv.org/abs/0801.3456}{{\ttfamily 0801.3456}}].

\bibitem{Curtin:2014cca}
D.~Curtin, R.~Essig, S.~Gori and J.~Shelton, \emph{{Illuminating Dark Photons with High-Energy Colliders}}, \href{http://dx.doi.org/10.1007/JHEP02(2015)157}{\emph{JHEP} {\bfseries 02} (2015) 157}, [\href{https://arxiv.org/abs/1412.0018}{{\ttfamily 1412.0018}}].

\bibitem{Aaij:2016qsm}
{\scshape $\rm LHCb$} collaboration, R.~Aaij et~al., \emph{{Search for long-lived scalar particles in $B^+ \to K^+ \chi (\mu^+\mu^-)$ decays}}, \href{http://dx.doi.org/10.1103/PhysRevD.95.071101}{\emph{Phys. Rev.} {\bfseries D95} (2017) 071101}, [\href{https://arxiv.org/abs/1612.07818}{{\ttfamily 1612.07818}}].

\bibitem{Aaij:2015tna}
{\scshape $\rm LHCb$} collaboration, R.~Aaij et~al., \emph{{Search for hidden-sector bosons in $B^0 \!\to K^{*0}\mu^+\mu^-$ decays}}, \href{http://dx.doi.org/10.1103/PhysRevLett.115.161802}{\emph{Phys. Rev. Lett.} {\bfseries 115} (2015) 161802}, [\href{https://arxiv.org/abs/1508.04094}{{\ttfamily 1508.04094}}].

\bibitem{CMS:2024bvl}
{\scshape CMS} collaboration, A.~Hayrapetyan et~al., \emph{{Search for long-lived particles decaying in the CMS muon detectors in proton-proton collisions at s=13\,\,TeV}}, \href{http://dx.doi.org/10.1103/PhysRevD.110.032007}{\emph{Phys. Rev. D} {\bfseries 110} (2024) 032007}, [\href{https://arxiv.org/abs/2402.01898}{{\ttfamily 2402.01898}}].

\bibitem{CMS:2024qxz}
{\scshape CMS} collaboration, A.~Hayrapetyan et~al., \emph{{Search for long-lived particles decaying to final states with a pair of muons in proton-proton collisions at $ \sqrt{s} $ = 13.6 TeV}}, \href{http://dx.doi.org/10.1007/JHEP05(2024)047}{\emph{JHEP} {\bfseries 05} (2024) 047}, [\href{https://arxiv.org/abs/2402.14491}{{\ttfamily 2402.14491}}].

\bibitem{FASER:2024bbl}
{\scshape FASER} collaboration, R.~Mammen~Abraham et~al., \emph{{Shining light on the dark sector: search for axion-like particles and other new physics in photonic final states with FASER}}, \href{http://dx.doi.org/10.1007/JHEP01(2025)199}{\emph{JHEP} {\bfseries 01} (2025) 199}, [\href{https://arxiv.org/abs/2410.10363}{{\ttfamily 2410.10363}}].

\bibitem{Dercks:2018eua}
D.~Dercks, J.~De~Vries, H.~K. Dreiner and Z.~S. Wang, \emph{{R-parity Violation and Light Neutralinos at CODEX-b, FASER, and MATHUSLA}}, \href{http://dx.doi.org/10.1103/PhysRevD.99.055039}{\emph{Phys. Rev.} {\bfseries D99} (2019) 055039}, [\href{https://arxiv.org/abs/1810.03617}{{\ttfamily 1810.03617}}].

\bibitem{Helo:2018qej}
J.~C. Helo, M.~Hirsch and Z.~S. Wang, \emph{{Heavy neutral fermions at the high-luminosity LHC}}, \href{http://dx.doi.org/10.1007/JHEP07(2018)056}{\emph{JHEP} {\bfseries 07} (2018) 056}, [\href{https://arxiv.org/abs/1803.02212}{{\ttfamily 1803.02212}}].

\bibitem{Graham:2015cka}
P.~W. Graham, D.~E. Kaplan and S.~Rajendran, \emph{{Cosmological Relaxation of the Electroweak Scale}}, \href{http://dx.doi.org/10.1103/PhysRevLett.115.221801}{\emph{Phys. Rev. Lett.} {\bfseries 115} (2015) 221801}, [\href{https://arxiv.org/abs/1504.07551}{{\ttfamily 1504.07551}}].

\bibitem{Flacke:2016szy}
T.~Flacke, C.~Frugiuele, E.~Fuchs, R.~S. Gupta and G.~Perez, \emph{{Phenomenology of relaxion-Higgs mixing}}, \href{http://dx.doi.org/10.1007/JHEP06(2017)050}{\emph{JHEP} {\bfseries 06} (2017) 050}, [\href{https://arxiv.org/abs/1610.02025}{{\ttfamily 1610.02025}}].

\bibitem{Choi:2016luu}
K.~Choi and S.~H. Im, \emph{{Constraints on Relaxion Windows}}, \href{http://dx.doi.org/10.1007/JHEP12(2016)093}{\emph{JHEP} {\bfseries 12} (2016) 093}, [\href{https://arxiv.org/abs/1610.00680}{{\ttfamily 1610.00680}}].

\bibitem{Craig:2014aea}
N.~Craig, S.~Knapen and P.~Longhi, \emph{{Neutral Naturalness from Orbifold Higgs Models}}, \href{http://dx.doi.org/10.1103/PhysRevLett.114.061803}{\emph{Phys. Rev. Lett.} {\bfseries 114} (2015) 061803}, [\href{https://arxiv.org/abs/1410.6808}{{\ttfamily 1410.6808}}].

\bibitem{Craig:2015pha}
N.~Craig, A.~Katz, M.~Strassler and R.~Sundrum, \emph{{Naturalness in the Dark at the LHC}}, \href{http://dx.doi.org/10.1007/JHEP07(2015)105}{\emph{JHEP} {\bfseries 07} (2015) 105}, [\href{https://arxiv.org/abs/1501.05310}{{\ttfamily 1501.05310}}].

\bibitem{Chacko:2005pe}
Z.~Chacko, H.-S. Goh and R.~Harnik, \emph{{The Twin Higgs: Natural electroweak breaking from mirror symmetry}}, \href{http://dx.doi.org/10.1103/PhysRevLett.96.231802}{\emph{Phys. Rev. Lett.} {\bfseries 96} (2006) 231802}, [\href{https://arxiv.org/abs/hep-ph/0506256}{{\ttfamily hep-ph/0506256}}].

\bibitem{Chacko:2005un}
Z.~Chacko, H.-S. Goh and R.~Harnik, \emph{{A Twin Higgs model from left-right symmetry}}, \href{http://dx.doi.org/10.1088/1126-6708/2006/01/108}{\emph{JHEP} {\bfseries 01} (2006) 108}, [\href{https://arxiv.org/abs/hep-ph/0512088}{{\ttfamily hep-ph/0512088}}].

\bibitem{Craig:2016kue}
N.~Craig, S.~Knapen, P.~Longhi and M.~Strassler, \emph{{The Vector-like Twin Higgs}}, \href{http://dx.doi.org/10.1007/JHEP07(2016)002}{\emph{JHEP} {\bfseries 07} (2016) 002}, [\href{https://arxiv.org/abs/1601.07181}{{\ttfamily 1601.07181}}].

\bibitem{Tucker-Smith:2001myb}
D.~Tucker-Smith and N.~Weiner, \emph{{Inelastic dark matter}}, \href{http://dx.doi.org/10.1103/PhysRevD.64.043502}{\emph{Phys. Rev. D} {\bfseries 64} (2001) 043502}, [\href{https://arxiv.org/abs/hep-ph/0101138}{{\ttfamily hep-ph/0101138}}].

\bibitem{Izaguirre:2015zva}
E.~Izaguirre, G.~Krnjaic and B.~Shuve, \emph{{Discovering Inelastic Thermal-Relic Dark Matter at Colliders}}, \href{http://dx.doi.org/10.1103/PhysRevD.93.063523}{\emph{Phys. Rev. D} {\bfseries 93} (2016) 063523}, [\href{https://arxiv.org/abs/1508.03050}{{\ttfamily 1508.03050}}].

\bibitem{Berlin:2018jbm}
A.~Berlin and F.~Kling, \emph{{Inelastic Dark Matter at the LHC Lifetime Frontier: ATLAS, CMS, LHCb, CODEX-b, FASER, and MATHUSLA}}, \href{http://dx.doi.org/10.1103/PhysRevD.99.015021}{\emph{Phys. Rev.} {\bfseries D99} (2019) 015021}, [\href{https://arxiv.org/abs/1810.01879}{{\ttfamily 1810.01879}}].

\bibitem{DAgnolo:2018wcn}
R.~T. D'Agnolo, C.~Mondino, J.~T. Ruderman and P.-J. Wang, \emph{{Exponentially Light Dark Matter from Coannihilation}}, \href{http://dx.doi.org/10.1007/JHEP08(2018)079}{\emph{JHEP} {\bfseries 08} (2018) 079}, [\href{https://arxiv.org/abs/1803.02901}{{\ttfamily 1803.02901}}].

\bibitem{Kaplan:2009ag}
D.~E. Kaplan, M.~A. Luty and K.~M. Zurek, \emph{{Asymmetric Dark Matter}}, \href{http://dx.doi.org/10.1103/PhysRevD.79.115016}{\emph{Phys. Rev.} {\bfseries D79} (2009) 115016}, [\href{https://arxiv.org/abs/0901.4117}{{\ttfamily 0901.4117}}].

\bibitem{Kim:2013ivd}
I.-W. Kim and K.~M. Zurek, \emph{{Flavor and Collider Signatures of Asymmetric Dark Matter}}, \href{http://dx.doi.org/10.1103/PhysRevD.89.035008}{\emph{Phys. Rev.} {\bfseries D89} (2014) 035008}, [\href{https://arxiv.org/abs/1310.2617}{{\ttfamily 1310.2617}}].

\bibitem{Zurek:2013wia}
K.~M. Zurek, \emph{{Asymmetric Dark Matter: Theories, Signatures, and Constraints}}, \href{http://dx.doi.org/10.1016/j.physrep.2013.12.001}{\emph{Phys. Rept.} {\bfseries 537} (2014) 91--121}, [\href{https://arxiv.org/abs/1308.0338}{{\ttfamily 1308.0338}}].

\bibitem{Hochberg:2014dra}
Y.~Hochberg, E.~Kuflik, T.~Volansky and J.~G. Wacker, \emph{{Mechanism for Thermal Relic Dark Matter of Strongly Interacting Massive Particles}}, \href{http://dx.doi.org/10.1103/PhysRevLett.113.171301}{\emph{Phys. Rev. Lett.} {\bfseries 113} (2014) 171301}, [\href{https://arxiv.org/abs/1402.5143}{{\ttfamily 1402.5143}}].

\bibitem{Hochberg:2014kqa}
Y.~Hochberg, E.~Kuflik, H.~Murayama, T.~Volansky and J.~G. Wacker, \emph{{Model for Thermal Relic Dark Matter of Strongly Interacting Massive Particles}}, \href{http://dx.doi.org/10.1103/PhysRevLett.115.021301}{\emph{Phys. Rev. Lett.} {\bfseries 115} (2015) 021301}, [\href{https://arxiv.org/abs/1411.3727}{{\ttfamily 1411.3727}}].

\bibitem{Choi:2017zww}
S.-M. Choi, Y.~Hochberg, E.~Kuflik, H.~M. Lee, Y.~Mambrini, H.~Murayama et~al., \emph{{Vector SIMP dark matter}}, \href{http://dx.doi.org/10.1007/JHEP10(2017)162}{\emph{JHEP} {\bfseries 10} (2017) 162}, [\href{https://arxiv.org/abs/1707.01434}{{\ttfamily 1707.01434}}].

\bibitem{Choi:2018iit}
S.-M. Choi, H.~M. Lee, P.~Ko and A.~Natale, \emph{{Resolving phenomenological problems with strongly-interacting-massive-particle models with dark vector resonances}}, \href{http://dx.doi.org/10.1103/PhysRevD.98.015034}{\emph{Phys. Rev.} {\bfseries D98} (2018) 015034}, [\href{https://arxiv.org/abs/1801.07726}{{\ttfamily 1801.07726}}].

\bibitem{Hall:2009bx}
L.~J. Hall, K.~Jedamzik, J.~March-Russell and S.~M. West, \emph{{Freeze-In Production of FIMP Dark Matter}}, \href{http://dx.doi.org/10.1007/JHEP03(2010)080}{\emph{JHEP} {\bfseries 03} (2010) 080}, [\href{https://arxiv.org/abs/0911.1120}{{\ttfamily 0911.1120}}].

\bibitem{Cui:2012jh}
Y.~Cui and R.~Sundrum, \emph{{Baryogenesis for weakly interacting massive particles}}, \href{http://dx.doi.org/10.1103/PhysRevD.87.116013}{\emph{Phys. Rev.} {\bfseries D87} (2013) 116013}, [\href{https://arxiv.org/abs/1212.2973}{{\ttfamily 1212.2973}}].

\bibitem{Cui:2014twa}
Y.~Cui and B.~Shuve, \emph{{Probing Baryogenesis with Displaced Vertices at the LHC}}, \href{http://dx.doi.org/10.1007/JHEP02(2015)049}{\emph{JHEP} {\bfseries 02} (2015) 049}, [\href{https://arxiv.org/abs/1409.6729}{{\ttfamily 1409.6729}}].

\bibitem{McKeen:2015cuz}
D.~McKeen and A.~E. Nelson, \emph{{CP Violating Baryon Oscillations}}, \href{http://dx.doi.org/10.1103/PhysRevD.94.076002}{\emph{Phys. Rev.} {\bfseries D94} (2016) 076002}, [\href{https://arxiv.org/abs/1512.05359}{{\ttfamily 1512.05359}}].

\bibitem{Aitken:2017wie}
K.~Aitken, D.~McKeen, T.~Neder and A.~E. Nelson, \emph{{Baryogenesis from Oscillations of Charmed or Beautiful Baryons}}, \href{http://dx.doi.org/10.1103/PhysRevD.96.075009}{\emph{Phys. Rev.} {\bfseries D96} (2017) 075009}, [\href{https://arxiv.org/abs/1708.01259}{{\ttfamily 1708.01259}}].

\bibitem{Mohapatra:1986bd}
R.~N. Mohapatra and J.~W.~F. Valle, \emph{{Neutrino Mass and Baryon Number Nonconservation in Superstring Models}}, \href{http://dx.doi.org/10.1103/PhysRevD.34.1642}{\emph{Phys. Rev.} {\bfseries D34} (1986) 1642}.

\bibitem{Asaka:2005an}
T.~Asaka, S.~Blanchet and M.~Shaposhnikov, \emph{{The nuMSM, dark matter and neutrino masses}}, \href{http://dx.doi.org/10.1016/j.physletb.2005.09.070}{\emph{Phys. Lett. B} {\bfseries 631} (2005) 151--156}, [\href{https://arxiv.org/abs/hep-ph/0503065}{{\ttfamily hep-ph/0503065}}].

\bibitem{Asaka:2005pn}
T.~Asaka and M.~Shaposhnikov, \emph{{The $\nu$MSM, dark matter and baryon asymmetry of the universe}}, \href{http://dx.doi.org/10.1016/j.physletb.2005.06.020}{\emph{Phys. Lett. B} {\bfseries 620} (2005) 17--26}, [\href{https://arxiv.org/abs/hep-ph/0505013}{{\ttfamily hep-ph/0505013}}].

\bibitem{Batell:2017cmf}
B.~Batell, T.~Han, D.~McKeen and B.~Shams Es~Haghi, \emph{{Thermal Dark Matter Through the Dirac Neutrino Portal}}, \href{http://dx.doi.org/10.1103/PhysRevD.97.075016}{\emph{Phys. Rev. D} {\bfseries 97} (2018) 075016}, [\href{https://arxiv.org/abs/1709.07001}{{\ttfamily 1709.07001}}].

\bibitem{Greljo:2018ogz}
A.~Greljo, D.~J. Robinson, B.~Shakya and J.~Zupan, \emph{{$R(D^{(*)})$ from $W^{\prime}$ and right-handed neutrinos}}, \href{http://dx.doi.org/10.1007/JHEP09(2018)169}{\emph{JHEP} {\bfseries 09} (2018) 169}, [\href{https://arxiv.org/abs/1804.04642}{{\ttfamily 1804.04642}}].

\bibitem{Asadi:2018wea}
P.~Asadi, M.~R. Buckley and D.~Shih, \emph{{It's all right(-handed neutrinos): a new $W^{\prime}$ model for the $R(D^{(*)})$ anomaly}}, \href{http://dx.doi.org/10.1007/JHEP09(2018)010}{\emph{JHEP} {\bfseries 09} (2018) 010}, [\href{https://arxiv.org/abs/1804.04135}{{\ttfamily 1804.04135}}].

\bibitem{Robinson:2018gza}
D.~J. Robinson, B.~Shakya and J.~Zupan, \emph{{Right-handed neutrinos and $R(D^{(*)})$}}, \href{http://dx.doi.org/10.1007/JHEP02(2019)119}{\emph{JHEP} {\bfseries 02} (2019) 119}, [\href{https://arxiv.org/abs/1807.04753}{{\ttfamily 1807.04753}}].

\bibitem{Collaboration:2285580}
{\scshape ATLAS} collaboration, \emph{{Technical Design Report for the Phase-II Upgrade of the ATLAS Muon Spectrometer}},  Tech. Rep. CERN-LHCC-2017-017. ATLAS-TDR-026, CERN, Geneva, Sep, 2017.

\bibitem{Gorordo:2022rro}
T.~Gorordo, S.~Knapen, B.~Nachman, D.~J. Robinson and A.~Suresh, \emph{{Geometry optimization for long-lived particle detectors}}, \href{http://dx.doi.org/10.1088/1748-0221/18/09/P09012}{\emph{JINST} {\bfseries 18} (2023) P09012}, [\href{https://arxiv.org/abs/2211.08450}{{\ttfamily 2211.08450}}].

\bibitem{Agostinelli:2002hh}
{\scshape GEANT4} collaboration, S.~Agostinelli et~al., \emph{{GEANT4: A Simulation toolkit}}, \href{http://dx.doi.org/10.1016/S0168-9002(03)01368-8}{\emph{Nucl. Instrum. Meth.} {\bfseries A506} (2003) 250--303}.

\bibitem{CODEX-b:2024tdl}
{\scshape CODEX-$\rm b$} collaboration, G.~Aielli et~al., \emph{{Technical design report for the CODEX-$\beta$ demonstrator}},  \href{https://arxiv.org/abs/2406.12880}{{\ttfamily 2406.12880}}.

\bibitem{Proto:2024wjz}
{\scshape ATLAS Muon} collaboration, G.~Proto, \emph{{Study of environment friendly gas mixtures for the Resistive Plate Chambers of the ATLAS phase-2 upgrade}}, \href{http://dx.doi.org/10.1016/j.nima.2024.170014}{\emph{Nucl. Instrum. Meth. A} {\bfseries 1070} (2025) 170014}, [\href{https://arxiv.org/abs/2503.05615}{{\ttfamily 2503.05615}}].

\bibitem{Akiba:2018neu}
K.~C. Akiba et~al., \emph{{The HeRSCheL detector: high-rapidity shower counters for LHCb}}, \href{http://dx.doi.org/10.1088/1748-0221/13/04/P04017}{\emph{JINST} {\bfseries 13} (2018) P04017}, [\href{https://arxiv.org/abs/1801.04281}{{\ttfamily 1801.04281}}].

\bibitem{Massa:2716316}
{\scshape ATLAS Muon} collaboration, L.~Massa, \emph{{The BIS78 Resistive Plate Chambers upgrade of the ATLAS Muon Spectrometer for the LHC Run-3}}, \href{http://dx.doi.org/10.1088/1748-0221/15/10/C10026}{\emph{JINST} {\bfseries 15} (2020) C10026}, [\href{https://arxiv.org/abs/2004.12693}{{\ttfamily 2004.12693}}].

\bibitem{Aielli:2022awh}
G.~Aielli et~al., \emph{{The Road Ahead for CODEX-b}},  \href{https://arxiv.org/abs/2203.07316}{{\ttfamily 2203.07316}}.

\end{thebibliography}\endgroup

\appendix

\newpage
\section{CODEX-$\rm b$ Main documents}
The following are the main documents the \CODEXb collaboration has published.

\renewcommand{\arraystretch}{1.2}

\vspace{.5cm}
\begin{tabular}{|p{11cm}|p{6cm}|} \hline
    \textbf{paper} & \textbf{description} \\ \hline
\bibentry{Gligorov:2017nwh} & \begin{minipage}{\textwidth}
\vspace{1cm} Original \CODEXb proposal.    
\end{minipage} \\ \hline
 \bibentry{Aielli:2019ivi} & \begin{minipage}{\textwidth}
\vspace{.6cm} Expression of interest. \end{minipage} \\ \hline
\bibentry{Gorordo:2022rro} & \begin{minipage}{\textwidth}
\vspace{.8cm} Geometry optimization paper \end{minipage} \\ \hline
\bibentry{Aielli:2022awh} &  \begin{minipage}{\textwidth}
\vspace{.4cm} \CODEXb snowmass contribution. \end{minipage} \\ \hline
 \bibentry{CODEX-b:2024tdl} & \CODEXbeta TDR, detailing all details concerning its construction and installation. \\ \hline
\end{tabular}

\section{FTE estimation for software and LHCb integration}

\begin{enumerate}
\item \textbf{Integration with LHCb}: since we plan to use the LHCb triggers, we need to integrate \CODEXb with LHCb. Lessons from \CODEXbeta can be recycled, although a team of 2 experts should work 1.5 FTE/y. \textbf{This task requires 3 FTE/y.}
\item \textbf{\CODEXb ECS and DAQ}: together with the integration in LHCb, we need to develop our data acquisition and experiment control system, to process data from the RPC readout, ensure synchronicity with the LHC and LHCb, and control the experiment. \textbf{This task will need a team of 3 experts working 2 FTE/y, for a total of 6 FTE/y.}
\item \textbf{Simulation}: A full simulation package should be developed, with \textsc{Pythia}/\textsc{MadGraph} generators and a complete \textsc{Geant4} detector and shielding simulation. This will take a team of 3 people, at least 1 post-doc, working 0.5 years on the generation and 1.5 years on the simulation. Also, the fast simulation should be prepared for general public use, requiring 2 people working 0.5 FTE/y. \textbf{This is a total of 7 FTE/y.}
\item \textbf{Digitization}: this task should be straightforward and \textbf{will require 1 person working 0.5 FTE/y.}
\item \textbf{Reconstruction}: this task can be recycled from the work with \CODEXbeta, and \textbf{will require 3 people working 0.5 FTE/y, totaling 1.5 FTE/y.}
\item \textbf{Data analysis}: dedicated data analysis tools and automated DQM can be recycled from \CODEXbeta, and \textbf{will require 2 people working 0.5 FTE/y, totaling 1 FTE/y.}
\end{enumerate}
This leads to the summary of \cref{tab:personnel_backup}. 

\begin{table}[h]
  \caption{Personnel time estimate for \CODEXb software development and integration with LHCb.\label{tab:personnel_backup}}
  \begin{tabular}{|l|r|r|r|}
    \toprule
    \multicolumn{1}{|c|}{task} & \multicolumn{3}{c|}{time [FTE years]} \\
    & baseline & scenario 1 & scenario 2 \\
    \midrule
    integration with LHCb & 3 & 3 & 3 \\
    \CODEXb ECS and DAQ & 6 & 6 & 6 \\
    \midrule
    DAQ/ECS total & 9 & 9 & 9 \\
    \midrule
    simulation & 7 & 7 & 7 \\
    digitization & 0.5 & 0.5 & 0.5 \\
    reconstruction & 1.5 & 1.5 & 1.5 \\
    data analysis & 1 & 1 & 1 \\ 
    \midrule
    software total & 10 & 10 & 10 \\
    \midrule
    grand total & 19 & 19 & 19 \\
    \bottomrule
  \end{tabular}
\end{table}

\newpage
\section{Additional Plots}
\begin{figure*}[h]
  \centering
  \includegraphics[width = 0.43\textwidth]{figures/new/haa_5}
  \includegraphics[width = 0.43\textwidth]{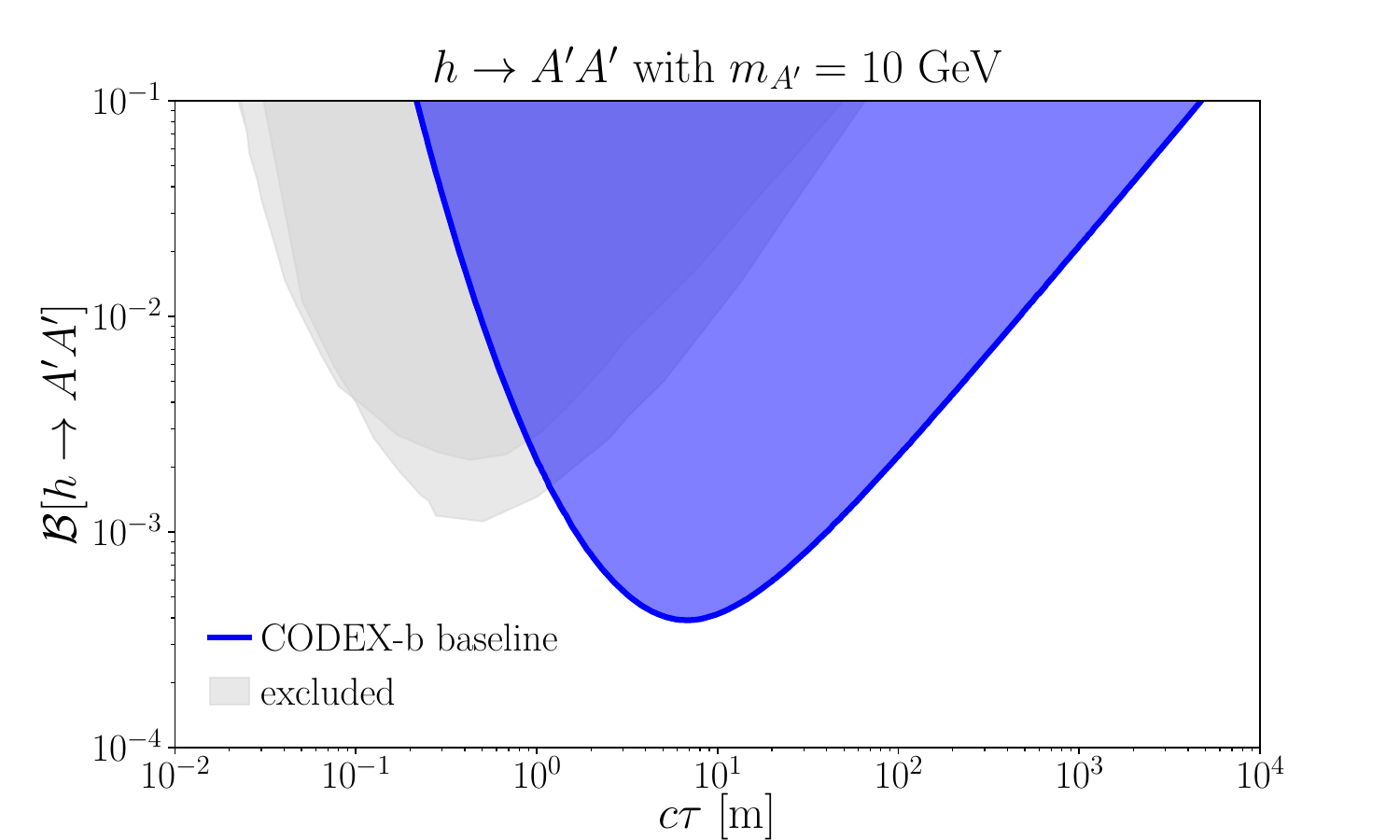}\\
  \includegraphics[width=0.43\textwidth]{figures/new/bh_0}
  \includegraphics[width=0.43\textwidth]{figures/new/bh_1}\\
  \includegraphics[width=0.43\textwidth]{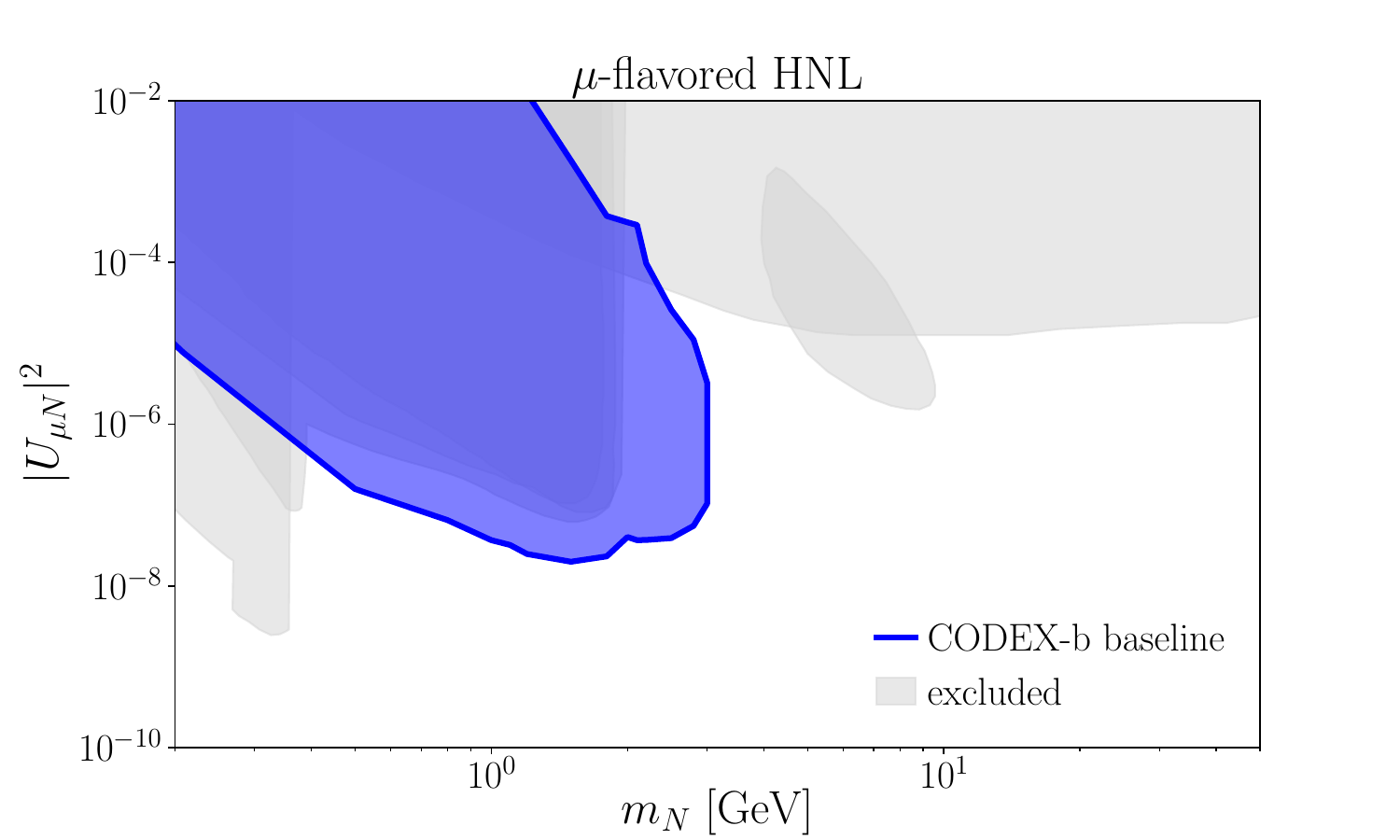}
  \includegraphics[width=0.43\textwidth]{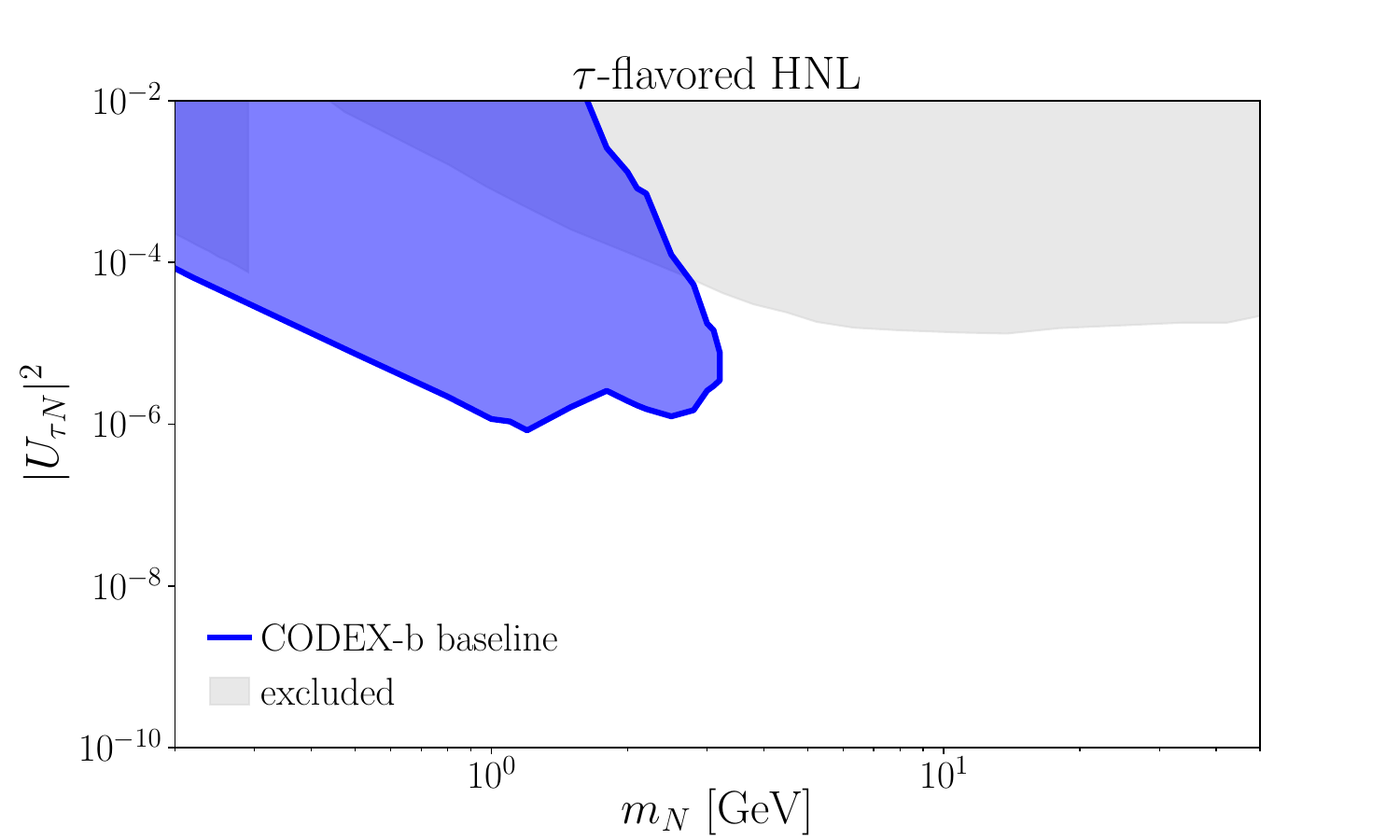}\\
  \includegraphics[width=0.43\textwidth]{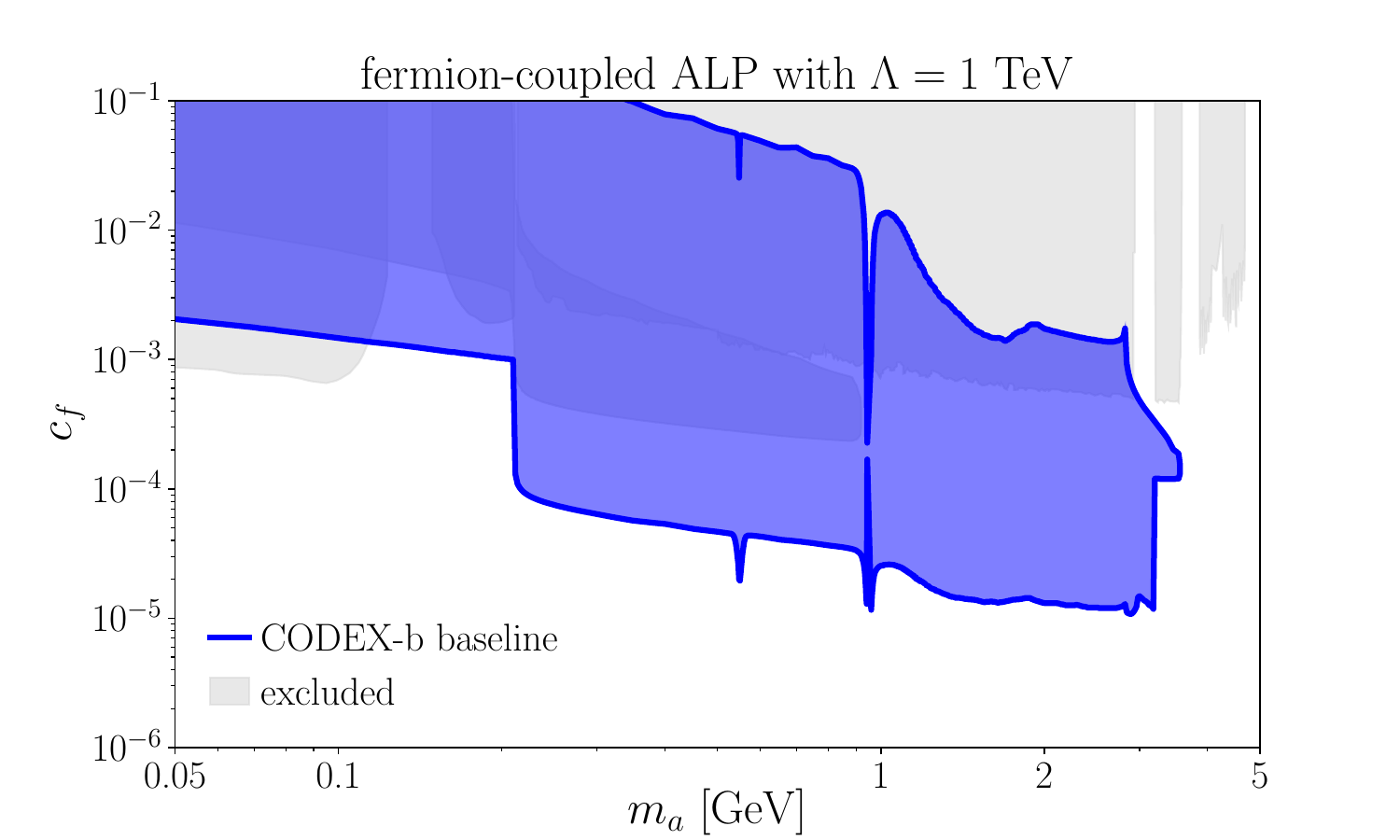}
  \includegraphics[width=0.43\textwidth]{figures/new/alp_g}\\
  \caption{\textbf{(top)} Baseline \CODEXb reach for $h\to A'A'$ with two different $A'$ masses compared to current exclusion limits. \textbf{(middle upper)} Simplified dark Higgs model reach with the mixed quartic coupling $\lambda_D$ chosen such that (left) $\mathcal{B}[h\to SS]=0$ and (right) $\mathcal{B}[h\to SS]=0.01$. \textbf{(middle lower)} Reach for (left) $\mu$- and (right) $\tau$-flavored HNLs. \textbf{(bottom)} Reach for (left) fermion- and (right) gluon-coupled ALPs. Plots modified from \ccite{Aielli:2019ivi} with updated exclusion limits~\cite{CMS:2021sch,CMS:2024bvl,CMS:2024qxz,FASER:2024bbl}.\label{fig:portals}}
\end{figure*}

\end{document}